\documentclass[twocolumn,pra,amsmath,amssymb,superscriptaddress,nofootinbib,longbibliography]{revtex4-2}

\usepackage{graphicx}
\usepackage{amsmath,amssymb}
\usepackage{bm}
\usepackage{xcolor}
\usepackage{algpseudocode}
\usepackage{algorithm}
\usepackage{float}
\usepackage{booktabs}
\usepackage{multirow}
\usepackage{hyperref}

\makeatletter
\@namedef{b@apsrev42Control}{{}{}}
\makeatother

\newcommand{\hvol}{\mathrm{HV}}

\newcommand{\ket}[1]{|#1\rangle}

\begin{document}

\title{Quantum-Enhanced Multi-Objective Optimization}

\author{Maolin Luo}
\thanks{These authors contributed equally to this work.}
\affiliation{Quantum Science Center of Guangdong-Hong Kong-Macao Greater Bay Area, Shenzhen, China}
\author{Jiapei Zhuang}
\thanks{These authors contributed equally to this work.}
\altaffiliation{Corresponding author: zhuangjiapei@quantumsc.cn}
\affiliation{Quantum Science Center of Guangdong-Hong Kong-Macao Greater Bay Area, Shenzhen, China}
\author{Zuoheng Zou}
\affiliation{Quantum Science Center of Guangdong-Hong Kong-Macao Greater Bay Area, Shenzhen, China}
\author{Man-Hong Yung}
\affiliation{Department of Physics, Southern University of Science and Technology, Shenzhen 518055, China}

\date{September 2, 2026}

\begin{abstract}
Multi-objective combinatorial optimization requires identifying Pareto-optimal trade-off solutions among conflicting objectives, often making it more demanding than its single-objective counterpart. Although quantum multi-objective optimization methods have begun to emerge, most existing quantum optimization workflows are still built around single-objective or fixed-scalarization settings. Building on existing weighted-sum QAOA approaches to quantum multi-objective optimization, we propose QEMOO, a quantum-enhanced multi-objective optimization framework that combines Pareto-based selection and warm-started QAOA sampling in a multi-round protocol under the same total shot budget. We further introduce a PBI-inspired adaptive direction-update scheme to improve coverage in strongly conflicting benchmark regimes. Across three benchmark stages, QEMOO improves Pareto-front hypervolume over the single-pass weighted-sum QAOA baseline under matched shot budgets, suggesting a practical route toward shot-efficient quantum-assisted multi-objective optimization and its future applications.
\end{abstract}

\maketitle

\begin{figure*}[!t]
\centering
\includegraphics[width=\textwidth]{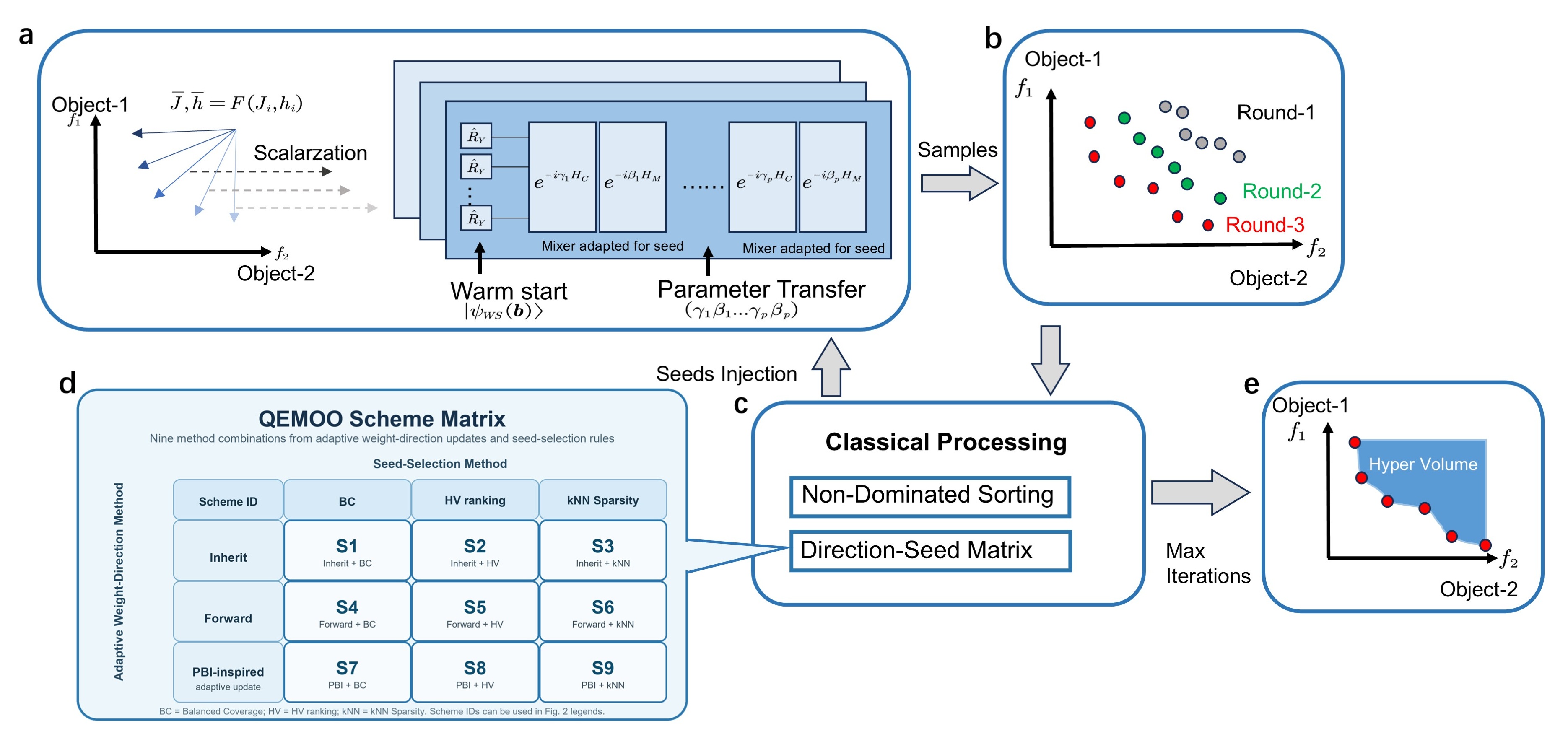}
\caption{Overview of the QEMOO framework. Panel~a shows how the multi-objective Ising model is scalarized into a pool of single-objective subproblems and solved by QAOA circuits with transferred parameters; after the first pass, selected elite bitstrings are injected into later rounds through seed-adapted warm-start mixers. Panel~b illustrates the round-wise samples in objective space, with Round-1, Round-2, and Round-3 shown in different colors. Panel~c shows the classical processing block: non-dominated sorting filters accumulated samples into elite candidates, and the Direction-Seed Matrix determines how adaptive weight-direction updates and seed-selection rules are combined for the next round. Panel~d gives the enlarged QEMOO Scheme Matrix, which enumerates the nine schemes S1--S9 by crossing three Adaptive Weight-Direction Methods (Inherit, Forward, and PBI-inspired) with three Seed-Selection Methods (BC, HV ranking, and kNN Sparsity), where BC denotes Balanced Coverage. Panel~e shows the final non-dominated set evaluated by the hypervolume indicator. The seed-injection arrow closes the feedback path from classical processing back to the warm-start QAOA circuits.}
\label{fig:overview}
\end{figure*}

\section{Introduction}
\label{sec:introduction}

Multi-objective combinatorial optimization (MOCO) problems arise pervasively across engineering design~\cite{ehrgott2005multicriteria}, logistics~\cite{heese2026supply_chain_logistics}, finance~\cite{aguilera2024portfolio}, and scientific discovery~\cite{xu2025multi}.
Many underlying single-objective problems are already NP-hard, and the multi-objective generalization can be harder even when the single-objective version admits efficient solutions~\cite{figueira2016easy}.
Classical multi-objective solvers such as NSGA-II~\cite{deb2002nsga2} and MOEA/D~\cite{zhang2007moea} have become de-facto standards, yet they rely on heuristic crossover and mutation operators that lack problem-specific guidance and can be computationally intensive for large combinatorial spaces.
Exact solvers for multi-objective integer programming~\cite{boland2017new, dachert2024simple} provide more accurate guarantees but become prohibitively time-consuming as problem size grows.
Quantum computing, with its ability to process global information via superposition and interference, can offers a candidate route to the generation of promising offspring and exploit sampling diversity to accelerate coverage of the Pareto front.

Quantum computing offers a fundamentally different paradigm that may provide advantages for single- and multi-objective combinatorial optimization. A central example is the Quantum Approximate Optimization Algorithm (QAOA)~\cite{farhi2014qaoa, harrigan2021qaoa}, introduced by Farhi, Goldstone, and Gutmann, which approximates low-energy states by alternating problem-Hamiltonian phase operators and mixing operators. QAOA has been studied extensively, both numerically and theoretically, as a potentially advantageous approach for hard optimization tasks~\cite{farhi2022qaoa_sk,boulebnane2025,Lykov_2023,Shaydulin_2024}. To reduce the cost of variational parameter optimization, we build on parameter-transfer techniques~\cite{sureshbabu2024parameter, shaydulin2023parameter_transfer}, which reuse correlated QAOA angles across related instances. In addition, hybrid optimization algorithms that combine the exploratory capability of quantum circuits with classical search and post-processing have been explored as a way to overcome the respective limitations of purely quantum or purely classical approaches~\cite{Zhu_2026,chandarana2025}. Despite substantial advances in single-objective optimization, extending these capabilities to multi-objective environments remains a critical frontier.

The Quantum Approximate Multi-Objective Optimization (QAMOO) framework~\cite{kotil2025qamoo} attempts this extension by decomposing MOCO problems into weighted-sum subproblems solved via QAOA. Despite this conceptual advance, QAMOO has two key limitations. First, fixed weighted-sum scalarization cannot systematically target unsupported or non-convex Pareto regions~\cite{das1997normal, zhang2007moea}. Second, its single-pass sampling strategy cannot use early discoveries to guide later sampling rounds.

Concurrently, a variational quantum multi-objective optimization (VQMO) approach was proposed~\cite{ekstrom2025vqmo} that employs the hypervolume indicator directly as the optimization objective, and subsequent work~\cite{ekstrom2026archiving} incorporates archiving and substitution mechanisms.
Also there are some researchs based on quantum annealing, including general multi-objective annealing protocols, scalarization-based QUBO solving, job-shop scheduling, and portfolio optimization~\cite{king2025moo_annealing,ayodele2023scalarisation_qubo,schworm2024jobshop_annealing,aguilera2024portfolio,sawamura2026jobshop_annealing}.

\begin{figure*}[!t]
    \centering
\includegraphics[width=\textwidth]{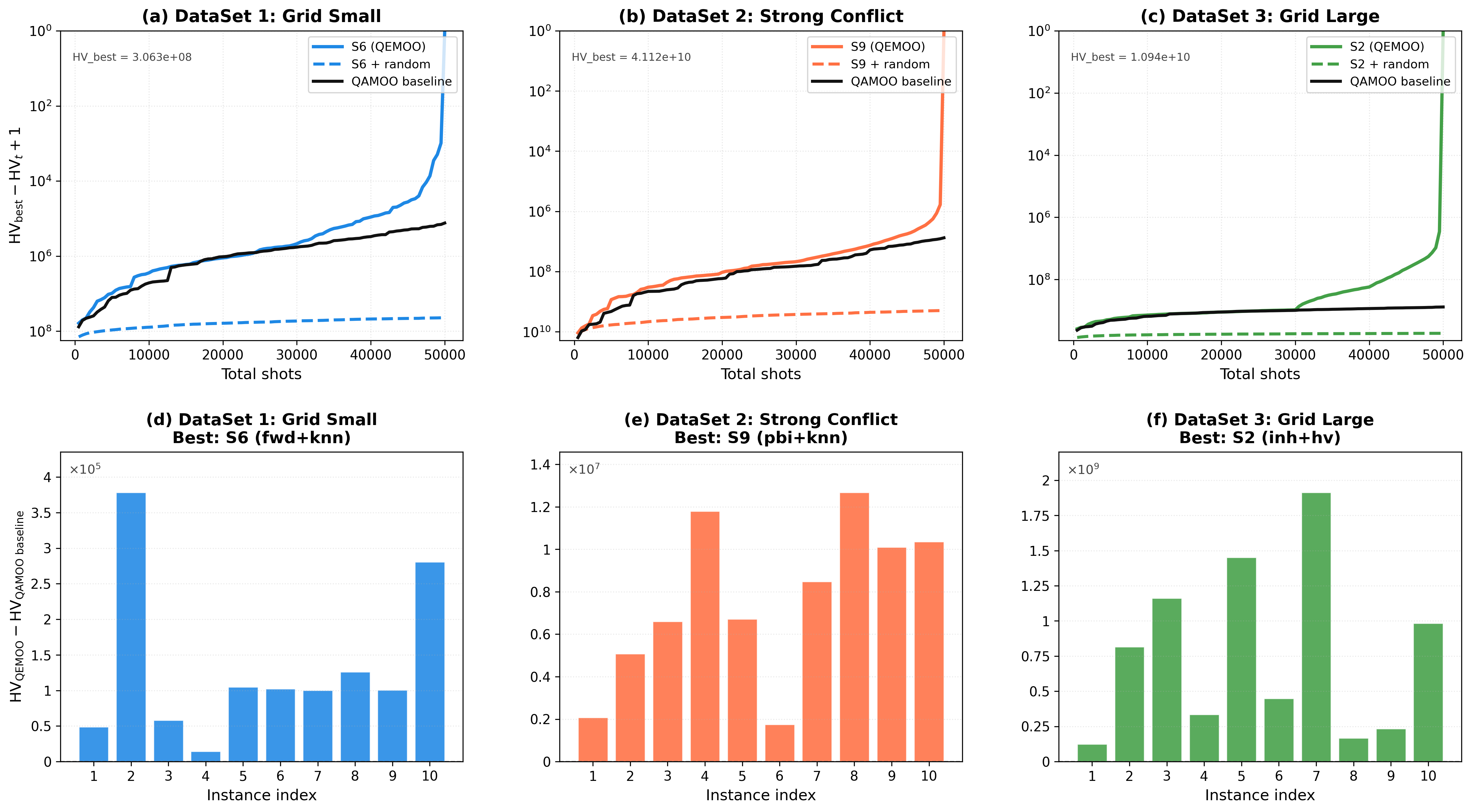}
\caption{Budget-normalized performance comparison and instance-level gains of the best QEMOO scheme on each benchmark dataset. The upper row reports HV-shots curves for DataSet~1: Grid Small, DataSet~2: Strong Conflict, and DataSet~3: Grid Large, respectively. Each upper-row panel compares the best QEMOO scheme for that dataset (S6, S9, or S2), the corresponding matched random-sampling variant (S6 + random, S9 + random, or S2 + random), and the single-pass QAMOO baseline of \citet{kotil2025qamoo}. Scheme IDs S1--S9 are defined by the QEMOO scheme matrix in Fig.~\ref{fig:overview}. The matched random variant keeps the same adaptive weight-direction and seed-selection configuration as the corresponding QEMOO scheme but replaces QAOA samples with uniformly random bitstrings under the same budget, isolating the contribution of quantum sampling. The upper-row vertical axis is $\mathrm{HV}_{\mathrm{best}}-\mathrm{HV}_{t}+1$, where $\mathrm{HV}_{\mathrm{best}}$ is the best observed mean HV within the same dataset and $\mathrm{HV}_{t}$ is the mean HV at cumulative shot count~$t$; the best value is plotted at the top, and lower positions indicate larger gaps from the best observed mean HV. The lower row reports the per-instance raw-HV gain of the best QEMOO scheme over the corresponding 1-round QAMOO baseline, so each bar shows the concrete improvement on one benchmark instance rather than an average over instances.}
    \label{fig:hv_shots_gap}
\end{figure*}

In this paper, we propose a \emph{Quantum-Enhanced Multi-Objective Optimization framework} for multi-objective combinatorial optimization to address the above limitations (see Fig.~\ref{fig:overview}).
Our main contributions are as follows:

\begin{enumerate}
    \item \textbf{Modular QEMOO framework.}
    We introduce a unified Quantum-Enhanced Multi-Objective Optimization (QEMOO) framework in which adaptive weight-direction updates and seed-selection strategies are treated as modular components that can be flexibly combined and systematically compared. This formulation enables a controlled study of how scalarization adaptation and elite transfer jointly affect Pareto-front discovery under a fixed quantum sampling budget.

    \item \textbf{Multi-round feedback allocation.}
    We propose a multi-round sampling protocol that redistributes a fixed total quantum budget across rounds while preserving transferred QAOA angles across problem instances and injecting warm-start state bias across rounds. This design converts intermediate non-dominated solutions into actionable feedback for later sampling rounds without increasing the total budget.

    \item \textbf{Evidences for quantum sampling gains.}
    Through HV-shots curves and matched random-sampling controls, we show that the quantum sampler contributes beyond the outer feedback logic alone (Fig.~\ref{fig:hv_shots_gap}).
\end{enumerate}

The remainder of this paper is organized as follows.
Section~\ref{sec:background} reviews the necessary background on multi-objective optimization, the Ising model, QAOA, and the QAMOO.
Section~\ref{sec:method} presents the Quantum-Enhanced Multi-Objective Optimization framework in detail.
Section~\ref{sec:experiments} describes the experimental setup, and Section~\ref{sec:results} reports the results and analysis.
We discuss the implications and limitations of our findings in Section~\ref{sec:discussion} and conclude in Section~\ref{sec:conclusion}.

\section{Background}
\label{sec:background}

\subsection{Multi-Objective Combinatorial Optimization}
\label{sec:bg:moo}

A multi-objective combinatorial optimization (MOCO) problem seeks to simultaneously minimize $k$ objective functions $f_1, f_2, \ldots, f_k$ over a finite decision space~$\mathcal{X}$~\cite{ehrgott2005multicriteria}:
\begin{equation}
\min_{\bm{x} \in \mathcal{X}} \; \bigl(f_1(\bm{x}),\, f_2(\bm{x}),\, \ldots,\, f_k(\bm{x})\bigr).
\label{eq:moop}
\end{equation}
When objectives conflict, no single solution simultaneously minimizes all of them.
Instead, the solution concept is \emph{Pareto optimality}: a solution~$\bm{x}$ \emph{dominates}~$\bm{x}'$ (written $\bm{x} \prec \bm{x}'$) if $f_i(\bm{x}) \leq f_i(\bm{x}')$ for all $i \in \{1,\ldots,k\}$ and $f_j(\bm{x}) < f_j(\bm{x}')$ for at least one~$j$.
A solution is \emph{Pareto optimal} if no other feasible solution dominates it.
The set of all Pareto-optimal objective vectors forms the \emph{Pareto front}~$\mathcal{F}^*$.
In practice, algorithms return a finite approximation set~$\mathcal{A} \subseteq \mathbb{R}^k$ whose quality must be measured against~$\mathcal{F}^*$.

A widely used quality indicator is the \emph{hypervolume} (HV)~\cite{zitzler2003assessment,fonseca2006dimension_sweep,riquelme2015metrics}, defined as the Lebesgue measure of the region dominated by the approximation set~$\mathcal{A}$ and bounded above by a reference point~$\bm{r} \in \mathbb{R}^k$. The computational complexity of HV and advanced algorithms for its calculation are detailed in~\cite{beume2009hypervolume,guerreiro2022hypervolume}:
\begin{equation}
\hvol(\mathcal{A}, \bm{r}) \;=\; \Lambda\Bigl(\bigcup_{\bm a \in \mathcal A} [ \bm a, \bm r ]\Bigr),
\label{eq:hv}
\end{equation}
where $[\bm{a}, \bm{r}]$ denotes the axis-aligned hyper-rectangle between~$\bm{a}$ and~$\bm{r}$.
Higher HV values indicate a better (more extensive and well-distributed) approximation of the Pareto front.

A classical approach to generate diverse Pareto-optimal solutions is objective scalarization~\cite{ehrgott2005multicriteria,deb2002nsga2}.
The most straightforward method is the \emph{weighted-sum} decomposition.
For a weight vector $\bm{\lambda} = (\lambda_1, \ldots, \lambda_k) \in \Delta^{k-1}$, one solves:
\begin{equation}
\min_{\bm{x} \in \mathcal{X}} \; g^{\text{WS}}(\bm{x} \mid \bm{\lambda}) = \sum_{i=1}^{k} \lambda_i \, f_i(\bm{x}).
\label{eq:weighted_sum}
\end{equation}
However, standard weighted-sum scalarization can miss unsupported Pareto-optimal regions in the usual geometric interpretation of Pareto fronts.
To overcome this, the \emph{Penalty-based Boundary Intersection (PBI)} method~\cite{zhang2007moea} evaluates the objective vector $\bm{f}(\bm{x})$ relative to a reference ideal point $\bm{z}^*$. It minimizes a scalarized objective combining the projection along the weight vector~$\bm{\lambda}$ (distance $d_1$) and the perpendicular deviation (distance $d_2$):
\begin{equation}
\min_{\bm{x} \in \mathcal{X}} \; g^{\text{PBI}}(\bm{x} \mid \bm{\lambda}, \theta_p) = d_1 + \theta_p d_2,
\label{eq:pbi}
\end{equation}
where the convergence and diversity distances are defined as:
\begin{subequations}
\begin{align}
d_1 &= \frac{\|(\bm{f}(\bm{x}) - \bm{z}^*)^\top \bm{\lambda}\|}{\|\bm{\lambda}\|}, \\
d_2 &= \|\bm{f}(\bm{x}) - (\bm{z}^* + d_1 \tfrac{\bm{\lambda}}{\|\bm{\lambda}\|})\|.
\end{align}
\label{eq:pbi_distances}
\end{subequations}
Here, $\theta_p > 0$ is a penalty parameter balancing front convergence against solution sparsity.
By sweeping over a collection of weight vectors $\Lambda = \{\bm{\lambda}^{(1)}, \ldots, \bm{\lambda}^{(n_w)}\}$, these decomposition methods yield solutions spanning the entire Pareto front, with PBI providing a mechanism to target such unsupported regions through projection and perpendicular-deviation terms. A technical derivation of the local linearization underlying our PBI-inspired adaptive scalarization is detailed in Appendix~\ref{app:pbi_derivation}.

\subsection{Ising Model for Multi-Objective Optimization}
\label{sec:bg:ising}

Many combinatorial optimization problems can be mapped onto Ising spin-glass Hamiltonians~\cite{lucas2014ising}.
In the multi-objective setting considered here, we have $k$~separate Ising Hamiltonians sharing the same graph topology but with distinct coupling and field parameters.
Specifically, given a graph $G = (V, E)$ with $|V| = n$ vertices and $|E| = m$ edges, the $i$-th objective is defined by the Ising energy function
\begin{equation}
f_i(\bm{\sigma}) \;=\; \sum_{(u,v) \in E} J^{(i)}_{uv}\, \sigma_u \sigma_v \;+\; \sum_{v \in V} h^{(i)}_v\, \sigma_v,
\label{eq:ising_obj}
\end{equation}
where $\bm{\sigma} = (\sigma_1, \ldots, \sigma_n) \in \{+1, -1\}^n$ is a spin configuration, $J^{(i)}_{uv}$ are the edge couplings, and $h^{(i)}_v$ are the local fields for objective~$i$.

Under objective scalarization with weight $\bm{\lambda}$, the scalarized objective becomes a single Ising Hamiltonian with projected coefficients. For the basic weighted-sum approach:
\begin{equation}
f_{\bm{\lambda}}(\bm{\sigma}) \;=\; \sum_{(u,v) \in E} \bar{J}_{uv}\, \sigma_u \sigma_v \;+\; \sum_{v \in V} \bar{h}_v\, \sigma_v\,,
\label{eq:ising_projected}
\end{equation}
where $\bar{J}_{uv} = \sum_{i=1}^{k} \lambda_i\, J^{(i)}_{uv}$ and $\bar{h}_v = \sum_{i=1}^{k} \lambda_i\, h^{(i)}_v$.
For PBI, the scalarization is non-linear and therefore cannot be projected by the same direct coefficient aggregation used for weighted sums. In QEMOO, we address this through a local linearization that produces a quadratic surrogate Hamiltonian compatible with QAOA; its operational role in the adaptive weight-direction update is described in Section~\ref{sec:method:line}, while the full derivation is deferred to Appendix~\ref{app:pbi_derivation}.

\subsection{Quantum Approximate Optimization Algorithm}
\label{sec:bg:qaoa}

The Quantum Approximate Optimization Algorithm (QAOA)~\cite{farhi2014qaoa,farhi2015qaoa_bounded} prepares a parameterized quantum state by alternating $p$~layers of a \emph{cost unitary} and a \emph{mixing unitary}.
For a cost Hamiltonian
$H_C = \sum_{(u,v)} \bar{J}_{uv}\, Z_u Z_v + \sum_v \bar{h}_v\, Z_v$
encoding the scalarized Ising objective~\eqref{eq:ising_projected}, the standard $p$-layer QAOA circuit prepares the state
\begin{equation}
\ket{\bm{\gamma}, \bm{\beta}} \;=\; \prod_{\ell=1}^{p}\, e^{-i \beta_\ell H_M}\, e^{-i \gamma_\ell H_C}\; \ket{+}^{\otimes n},
\label{eq:qaoa_state}
\end{equation}
where $\ket{+}^{\otimes n}$ is the uniform superposition, $\bm{\gamma} = (\gamma_1, \ldots, \gamma_p)$ and $\bm{\beta} = (\beta_1, \ldots, \beta_p)$ are variational parameters, and the mixing Hamiltonian is $H_M = \sum_v X_v$.
In the standard QAOA, the initial state is the uniform superposition; in our warm-start variant, this is replaced by a product state biased towards an elite solution extracted from a previous search round (see Section~\ref{sec:method}).
At the circuit level, $e^{-i\gamma_\ell H_C}$ is decomposed into single-qubit $R_Z$ rotations for the field terms and two-qubit $R_{ZZ}$ rotations for the coupling terms, while $e^{-i\beta_\ell H_M}$ consists of $R_X$ rotations on each qubit~\cite{farhi2014qaoa}.
For weighted Ising instances, the $\gamma$ parameters are rescaled using a degree-dependent factor: $\gamma_\ell^{\text{eff}} = \gamma_\ell \cdot \arctan(1/\sqrt{D-1}) / s$, where $D$ is the average vertex degree and $s$ is the root-mean-square of the Ising coefficients~\cite{sureshbabu2024parameter}.

\textbf{Parameter transfer.}
A key enabling technique for our framework is \emph{parameter transfer}: near-optimal QAOA angles $(\bm{\gamma}^*, \bm{\beta}^*)$ are known to concentrate across instances drawn from the same problem class~\cite{farhi2022qaoa_sk,akshay2021parameter_concentrations,wurtz2021fixed_angle}.
For weighted MaxCut-type Ising problems, it has been demonstrated~\cite{sureshbabu2024parameter} that parameters optimized on small instances ($q_{\text{target}} = 2$ qubits) transfer effectively to larger ones, provided appropriate rescaling is applied.
This eliminates the classical outer loop that would otherwise dominate the computational cost of QAOA, making it feasible to run QAOA on hundreds of weight directions without per-direction parameter optimization~\cite{shaydulin2023parameter_transfer,montanez2024transfer,galda2021transferability}.
In this work, we use pre-computed transferred parameters at depth $p=3$ with $q_{\text{target}}=2$, following the protocol of~\cite{sureshbabu2024parameter}.

\subsection{The QAMOO}
\label{sec:bg:qamoo}

The Quantum Approximate Multi-Objective Optimization (QAMOO) algorithm~\cite{kotil2025qamoo} combines the weighted-sum decomposition~\eqref{eq:weighted_sum} with QAOA~\eqref{eq:qaoa_state} to approximate Pareto fronts for multi-objective Ising problems.
The procedure works as follows:

\begin{enumerate}
\item \emph{Weight generation.}
A set of $n_w$~weight vectors $\Lambda = \{\bm{\lambda}^{(1)}, \ldots, \bm{\lambda}^{(n_w)}\}$ is drawn from the $(k{-}1)$-dimensional probability simplex~$\Delta^{k-1}$. While traditional QAMOO utilizes uniform Dirichlet sampling, our framework explores structured distributions like simplex lattices and center-biased variants to optimize front coverage (detailed in Appendix~\ref{app:weights}).

\item \emph{QAOA sampling.}
For each weight~$\bm{\lambda}^{(j)}$, the projected Ising coefficients $(\bar{J}, \bar{h})$ are computed via~\eqref{eq:ising_projected}, a QAOA circuit is constructed using transferred parameters, and $s$~shots are sampled to produce spin configurations $\{\bm{\sigma}^{(j)}_1, \ldots, \bm{\sigma}^{(j)}_s\}$.

\item \emph{Non-dominated filtering.}
All sampled configurations across all directions are evaluated on all $k$~objectives, and non-dominated solutions are retained to form the approximation set~$\mathcal{A}$.

\item \emph{Evaluation.}
The hypervolume $\hvol(\mathcal{A}, \bm{r})$ is computed with respect to a fixed reference point~$\bm{r}$.
\end{enumerate}

In its original formulation, QAMOO performs a \emph{single pass}: each weight direction receives the same number of shots~$s$, and no information flows between different weight directions or across sequential sampling rounds.
The total quantum sampling budget is $B = n_w \times s$.
Our contribution is to introduce a multi-round extension that keeps the same total budget~$B$ but redistributes it across rounds with feedback from intermediate non-dominated solutions.

\section{Method}
\label{sec:method}

\subsection{Framework Overview}
\label{sec:method:overview}

We propose a Quantum-Enhanced Multi-Objective Optimization framework that extends the single-pass QAMOO pipeline (Section~\ref{sec:bg:qamoo}) into a multi-round iterative scheme capable of adapting the scalarized search directions across rounds under a fixed sampling budget.
The total quantum sampling budget $B = n_w \times s_{\text{total}}$ remains identical to the baseline; the key difference is that the per-weight shots are distributed across $R$~rounds with allocations $s_1, s_2, \ldots, s_R$ satisfying $\sum_{r=1}^{R} s_r = s_{\text{total}}$ (budget scaling analysis in Appendix~\ref{app:budget}).
In our default configuration, $R=3$ with a $3{:}1{:}1$ per-weight shot allocation, yielding $(s_1, s_2, s_3) = (300, 100, 100)$ and $s_{\text{total}}=500$ per weight. This front-loaded schedule ($s_1 > s_2, s_3$) ensures a broad, unbiased initial exploration of the objective space, while reserving sufficient budget for targeted exploitation of the discovered elites in subsequent rounds.
As shown in Fig.~\ref{fig:overview}(c), the classical feedback step is decomposed into two explicit operations. First, non-dominated sorting evaluates the accumulated samples from the current and previous rounds and extracts elite candidates in objective space. Second, the Direction-Seed Matrix assigns how those elites are reused by pairing an \emph{Adaptive Weight-Direction Method}, which determines how each weight direction is translated into a scalarized search target, with a \emph{Seed-Selection Method}, which determines which non-dominated elites are transferred across rounds. The enlarged QEMOO Scheme Matrix in Fig.~\ref{fig:overview}(d) summarizes these pairings: crossing the three direction-update rules (Inherit, Forward, and PBI-inspired) with the three seed-selection rules (BC, HV ranking, and kNN Sparsity) defines nine QEMOO schemes, denoted S1--S9, where BC denotes Balanced Coverage. All circuit calls share the same transferred angle template $(\bm{\gamma}^*, \bm{\beta}^*)$, while only the warm-start bias changes from round to round.

\begin{algorithm}[H]
\caption{QEMOO Multi-Round Optimization Procedure}
\label{alg:framework}
\begin{algorithmic}[1]
\Require Weight pool $\Lambda = \{\bm{\lambda}^{(j)}\}_{j=1}^{n_w}$, adaptive weight-direction rule $m_{\mathrm{awd}}$, shot schedule $(s_1, \ldots, s_R)$, warm-start coefficient $c$, transferred parameters $(\bm{\gamma}^*, \bm{\beta}^*)$
\Ensure Approximation set $\mathcal{A}$
\State $\mathcal{A} \leftarrow \emptyset$; \,\textbf{seeds} $\leftarrow [\text{None}]^{n_w}$
\For{$r = 1, 2, \ldots, R$}
    \For{$j = 1, 2, \ldots, n_w$}
        \State Construct the scalarized objective for $\bm{\lambda}^{(j)}$ according to $m_{\mathrm{awd}}$
        \If{$r = 1$}
            \State Build standard QAOA circuit with $\ket{+}^{\otimes n}$ init
        \Else
            \State Build warm-start circuit with seed $\bm{b}^{(j)}$ and coeff.~$c$
        \EndIf
        \State Sample $s_r$ shots $\rightarrow \{\bm{\sigma}^{(j)}_1, \ldots, \bm{\sigma}^{(j)}_{s_r}\}$
    \EndFor
    \State Evaluate all samples on $k$ objectives
    \State Update $\mathcal{A}$ with non-dominated solutions
    \If{$r < R$}
        \State $\textbf{seeds} \leftarrow \textsc{SelectEliteSeeds}(\mathcal{A}_r, n_w)$
        \Comment{Section~\ref{sec:method:seeds}}
    \EndIf
\EndFor
\State \Return $\mathcal{A}$
\end{algorithmic}
\end{algorithm}

The algorithm proceeds as follows (see also Algorithm~\ref{alg:framework} and Fig.~\ref{fig:overview}). Operationally, transferred angles remain fixed across rounds, whereas warm-starting injects the elite bias that distinguishes later rounds from the cold start; the colored Round-1, Round-2, and Round-3 sample clouds in Fig.~\ref{fig:overview}(b) visualize this iterative feedback process:

\begin{enumerate}
\item \textbf{Round~1 (Cold start).}
For each of the $n_w$~weight directions, construct a standard QAOA circuit with the transferred parameter template and initial state $\ket{+}^{\otimes n}$, structured to minimize either the standard weighted-sum target or the effective scalarized target induced by the PBI-inspired update, and sample $s_1$~shots.
Evaluate all samples on every objective and perform non-dominated sorting.

\item \textbf{Seed selection.}
From the non-dominated set, select $n_w$~elite seeds using target-oriented selectors (Section~\ref{sec:method:seeds}), and use the Direction-Seed Matrix to assign each selected seed to the corresponding adaptive weight direction.

\item \textbf{Round~$r > 1$ (Warm start).}
For each weight direction, construct a warm-start QAOA circuit (Section~\ref{sec:method:warmstart}) using the same transferred parameter template but initialized with the assigned seed solution, and sample $s_r$~shots.
Merge the new samples with all previous non-dominated solutions.

\item Repeat Steps~2--3 until all $R$ rounds are complete.
\end{enumerate}

This design introduces a \emph{feedback loop}: each round's non-dominated front informs the next round's initial quantum states, enabling adaptive reallocation of quantum resources toward the most promising regions of objective space.

\subsection{Adaptive Weight-Direction Strategies}
\label{sec:method:line}

The first design axis of QEMOO is the \emph{Adaptive Weight-Direction Method}, namely how a weight direction $\bm{\lambda}$ is converted into the scalarized optimization target solved by the QAOA circuit in that round. In the current framework, we consider three stage-aware method families. \texttt{Inherit} keeps the parent-associated weighted-sum direction unchanged, i.e., it reuses the inherited weight vector $\bm{\lambda}$ itself as the scalarization direction. \texttt{Forward} still uses weighted-sum scalarization, but it first deforms the inherited direction in weight space toward the one-hot anchor of the parent seed's strongest normalized objective dimension, yielding a projected direction $\bm{\lambda}^{\mathrm{fwd}} = \Pi_{\Delta}((1-\eta)\bm{\lambda} + \eta \bm{e}_{d^*})$. \texttt{PBI} is our shorthand for the PBI-inspired adaptive scalarization: it uses a local linearization of the non-linear PBI signal around the current elite solution to synthesize an effective direction for the next QAOA round. In this sense, \texttt{PBI} differs from \texttt{Inherit} and \texttt{Forward} at the direction-generation level rather than by requiring a different circuit-level Hamiltonian family; the synthesized direction is still implemented as a weighted-sum-like quadratic Ising Hamiltonian. This gives QEMOO a geometry-aware update that is particularly useful for the Strong Conflict benchmark, where fixed or inherited directions can miss important trade-off regions.

These adaptive weight-direction methods are not expected to be uniformly optimal across all landscapes. Instead, they form the geometry-facing axis of the framework, while the Seed-Selection Methods in the next subsection form the transfer-facing axis. The rerun benchmark therefore evaluates stage-appropriate combinations of adaptive weight-direction methods and seed-selection methods rather than imposing a single universal recipe.

\subsection{Target-Oriented Seed Selection Dynamics}
\label{sec:method:seeds}

The transition between rounds requires selecting $n_w$~elite seeds from the current round's non-dominated set.
A naive strategy---e.g., random selection or top-$n_w$ by a scalar metric---would fail to preserve diversity along the Pareto front.
We design a multi-criteria selection framework that supports three distinct targeting strategies aimed at maximizing the value of the subsequent warm-start operations:

\begin{itemize}
\item \textbf{Balanced-Coverage selector (BC):}
Inspired by NSGA-II~\cite{deb2002nsga2}, this selector balances extremity, front coverage, and controlled diversity.
First, anchor points (the extrema of each objective dimension) are unconditionally included.
Then, remaining candidates are prioritized by their crowding distance in normalized objective space.
To prevent over-clustering in dense regions, a progressive distance-threshold relaxation mechanism enforces a minimum Euclidean separation $\tau$ between seeds, guaranteeing broad coverage.

\item \textbf{Hypervolume-ranking selector (HV ranking):}
Instead of relying purely on spatial density, this selector iteratively adds the non-dominated point that yields the maximal marginal increase in hypervolume relative to the currently selected seed subset. This greedy approach directly aligns the seed selection with the ultimate evaluation metric.

\item \textbf{kNN-sparsity selector (kNN Sparsity):}
This selector ranks candidates by a local sparsity score derived from their $k$-nearest-neighbor spacing in normalized objective space, favoring seeds that remain informative while staying geometrically separated from already crowded frontier regions.

Directional cap restriction: Regardless of the active selector, we impose a cap of $d_{\max}$~seeds per weight direction (default $d_{\max}=2$). This constraint forces the algorithm to distribute warm-start biases uniformly across diverse regions of the weight simplex, preventing an artificial collapse of the Pareto front. Without this cap, a uniquely flat or easily optimized region of the objective space could dominate the non-dominated set, ``hijacking'' the quantum budget for the next round and starving complex trade-off zones.
\end{itemize}

\subsection{QAOA Circuit Instantiation: Transfer and Warm Start}
\label{sec:method:warmstart}

The actual QAOA circuit used in QEMOO is built from two layers of inductive bias: transferred angles across problem instances, and warm-start state bias across rounds. Section~\ref{sec:bg:qaoa} provides the theoretical basis for parameter transfer; here we only specify how the shared transferred template $(\bm{\gamma}^*, \bm{\beta}^*)$ is instantiated together with the round-dependent warm-start bias.

The standard QAOA circuit begins from the uniform superposition $\ket{+}^{\otimes n}$, which encodes no prior knowledge about the solution landscape.
Following the warm-starting paradigm introduced in~\cite{egger2021warmstart}, we replace this initial state with a \emph{biased} product state that encodes a previously discovered elite solution.

Given a seed bitstring $\bm{b} = (b_1, \ldots, b_n) \in \{0,1\}^n$ and a bias coefficient $c \in [0,1]$, the initial state of each qubit~$q$ is prepared as $R_Y(\theta_q) \ket{0}$, where the rotation angle is
\begin{equation}
\theta_q = 2 \arcsin\!\Bigl(\sqrt{(1 - c) \cdot \tfrac{1}{2} + c \cdot b_q}\Bigr).
\label{eq:warm_theta}
\end{equation}
This construction interpolates smoothly between two extremes:
\begin{itemize}
\item When $c = 0$: $\theta_q = 2\arcsin(1/\sqrt{2}) = \pi/2$ for all~$q$, recovering $\ket{+}^{\otimes n}$ (standard QAOA).
\item When $c \to 1$: $\theta_q \to \pi \cdot b_q$, preparing a state close to the computational basis state~$\ket{\bm{b}}$ (greedy exploitation).
\end{itemize}

The mixing Hamiltonian must also be adapted to preserve the biased initial state as a fixed point.
Following~\cite{egger2021warmstart}, we replace the standard $R_X(2\beta)$ mixer with a \emph{rotated} mixer:
\begin{equation}
U_M^{(\text{warm})}(\beta) = \prod_{q=1}^{n} R_Y(\theta_q)\, R_Z(2\beta)\, R_Y(-\theta_q),
\label{eq:warm_mixer}
\end{equation}
which applies a $Z$-rotation in the frame rotated by $\theta_q$.
This ensures that the mixer's fixed point aligns with the biased initial state rather than the uniform superposition, so the warm-start bias is preserved throughout the circuit.
Importantly, compiling the rotated mixer requires only single-qubit $R_Y$ and $R_Z$ gates, which are natively available on superconducting and ion-trap architectures. Thus, the warm-start mechanism incurs negligible additional circuit depth or hardware overhead compared to standard QAOA.

The complete warm-start QAOA circuit for layer~$\ell$ is thus:
\begin{equation}
U_\ell = U_M^{(\text{warm})}(\beta_\ell) \cdot U_C(\gamma_\ell^{\text{eff}}),
\label{eq:warm_layer}
\end{equation}
where $U_C(\gamma_\ell^{\text{eff}})$ is the same cost unitary as in standard QAOA (composed of $R_Z$ and $R_{ZZ}$ gates), and $\gamma_\ell^{\text{eff}}$ is the rescaled $\gamma$ parameter using the degree-dependent formula described in Section~\ref{sec:bg:qaoa}.

\textbf{Choice of $c$.}
The bias coefficient $c$ controls the exploration--exploitation trade-off.
Too small a~$c$ renders the warm-start ineffective, while too large a~$c$ causes the search to collapse around the seed solution, preventing discovery of new Pareto-optimal points.
We determine the optimal $c$ through systematic parameter scanning in Section~\ref{sec:results}.

\section{Experimental Setup}
\label{sec:experiments}

\subsection{Benchmark Problems}
\label{sec:exp:instances}

We evaluate our method on a diverse benchmark suite of multi-objective Ising combinatorial optimization instances. The objectives exhibit significant mutual conflict, ensuring a non-trivial Pareto front.
\begin{itemize}
\item \textbf{Grid Small (grid4x5):} Ten public 5-objective Ising instances defined on a $4 \times 5$ grid ($n=20$ spins, $m=31$ edges). The couplings are generated from shared case-level random coefficient profiles with moderate edge and field scales, producing moderate-conflict random-grid benchmark instances. This stage is evaluated with the state-vector backend \texttt{mqvector}.
\item \textbf{Grid Large (grid6x7):} Ten public 5-objective instances defined on a $6 \times 7$ grid ($n=42$ spins, $m=71$ edges). They follow the same random-grid construction as the small suite but at a larger size, and are treated as the large-scale tensor-network stage. This benchmark is evaluated with the tensor-network backend \texttt{mqmps}; the latest rerun results reported in the main text use bond dimension $\chi=30$.
\item \textbf{Strong Conflict:} Ten engineered benchmark instances in which objective-asymmetric edge couplings make one objective strongly favor an edge while the others penalize it, intentionally producing sharply conflicting objective preferences and a highly warped trade-off structure. This stage uses $k=4$, $n=18$, and $m=60$, and is evaluated with \texttt{mqvector} as the dedicated stress test for the PBI-inspired adaptive scalarization method.
\end{itemize}

Across all three stages, we keep the default weight-pool protocol fixed so that the benchmark differences are driven by problem topology and method design rather than by changing directional coverage. Since that pool is part of the experimental protocol rather than of the dataset definition itself, we specify it in Sec.~\ref{sec:exp:impl}.

\subsection{Algorithm Configurations and Method Combinations}
\label{sec:exp:methods}

We compare the solvers by varying the \emph{Adaptive Weight-Direction Methods} (how scalarization is performed) and the \emph{Seed-Selection Methods} (how warm-starts are selected), all operating under an identical total sampling budget of $B_{\text{total}} = 50,000$ shots per instance. This ensures that performance gains stem from the adaptive feedback loop rather than increased query number.

\begin{itemize}
\item \textbf{Quantum Baseline:}
The non-iterative weighted-sum QAOA approach~\cite{kotil2025qamoo}. For each of the $n_w$ weight directions, run the single-round QAOA circuit with $s_{\text{total}}$ shots.

\item \textbf{QEMOO:}
Our iterative method operates in $R = 3$ rounds according to shot budgets distributed in a ratio of $3{:}1{:}1$, with the same total budget $s_{\text{total}}$. To systematically investigate the QEMOO framework, we evaluate the method combinations identified by the S1--S9 scheme matrix in Fig.~\ref{fig:overview}. In the current rerun benchmark, the Grid Small and Grid Large stages use the Inherit/Forward rows (S1--S6), while the Strong Conflict stage uses the PBI-inspired row (S7--S9); all stages compare the same three seed-selection columns, BC (Balanced Coverage), HV ranking, and kNN Sparsity.

\item \textbf{Classical Baseline (Uniform Sampling):}
Uniformly random spin configuration sampling acts as a classical baseline.
\end{itemize}

For all QAOA evaluations, circuit depth is $p=3$ utilizing pre-optimized parameters transferred from $q_{\text{target}}=2$ small-scale analogues~\cite{sureshbabu2024parameter}.

For the weight-pool analysis experiments (Appendix~\ref{app:weights}), we additionally compare all nine weight-pool schemes under both baseline and warm-start configurations.
For the Grid-Large ablations in Appendix~\ref{app:large_mps_hyperparams}, we keep the multi-round protocol fixed and vary only one hyperparameter at a time, so those appendix studies should be read as controlled support for the main benchmark rather than as a separate evaluation track.

\subsection{Evaluation Metrics}
\label{sec:exp:metrics}

The Hypervolume indicator is computed in the \emph{raw Ising energy space} for each objective. To ensure consistent evaluation across different problem topologies, we define a case-specific reference point $\bm{r} = (B_1, \dots, B_k)$, where $B_i = \sum_{(u,v)} |J^{(i)}_{uv}| + \sum_v |h^{(i)}_v|$ represents a conservative bound on the maximum possible objective value for any spin configuration.
State-vector simulations for the Grid Small and Strong Conflict instances utilize the \texttt{mqvector} backend, while large-scale grid benchmarks leverage the tensor-network \texttt{mqmps} backend.

The primary metric is the hypervolume indicator $\hvol(\mathcal{A}, \bm{r})$~\eqref{eq:hv}, computed using the \texttt{pygmo} library~\cite{biscani2020pygmo}.

We also report: (i)~the mean HV gain $\Delta\hvol = \hvol_{\text{warm}} - \hvol_{\text{baseline}}$ averaged across the ten instances; (ii)~the number of \emph{positive cases} (instances where $\Delta\hvol > 0$); and (iii)~the number of non-dominated solutions retained in the final approximation set.

\subsection{Default Configuration and Controlled Ablations}
\label{sec:exp:impl}

All experiments are run on MindQuantum~\cite{mindquantum2021} using the state-vector simulator (\texttt{mqvector}) and matrix product state (MPS) backend (\texttt{mqmps}); the main Grid-Large rerun uses bond dimension $\chi = 30$. QAOA circuits use depth $p=3$ with parameters pre-computed via the transfer protocol described in~\cite{sureshbabu2024parameter} at $q_{\text{target}}=2$.

Unless noted otherwise, all stages share the same default directional-coverage protocol: the weight pool is drawn from a \texttt{dirichlet\_uniform} distribution (\texttt{Dirichlet}$(\alpha=1)$ on the simplex) with seed 2026, and each run uses the first $n_w = 100$ weight directions from that shared pool for both the baseline and the warm-start rounds. Non-dominated sorting is performed via PyGMO's fast non-dominated sorting~\cite{biscani2020pygmo}. All experiments use fixed random seeds to ensure full reproducibility, and single-threaded execution to avoid non-deterministic parallelism artifacts.
Whenever we move beyond this default setting in the appendix, we do so through one-factor-at-a-time scans: Appendix~\ref{app:weights} studies how the weight-pool geometry changes transfer behavior, Appendix~\ref{app:large_mps_hyperparams} studies the Grid-Large hyperparameters $(c,\chi,n_w)$, and Appendix~\ref{app:budget} studies shot-budget sensitivity. This separation is intentional: the main text reports the stage-wise benchmark outcomes, while the appendix explains the mechanisms and parameter choices that support those outcomes.

\section{Results and Analysis}
\label{sec:results}

In this section, we present the stage-wise evaluation of the QEMOO framework. The rerun benchmark compares stage-appropriate \textbf{Adaptive Weight-Direction Methods} (Inherit, Forward, or PBI, depending on the landscape) together with three \textbf{Seed-Selection Methods} (BC, HV ranking, kNN Sparsity). The main text follows a single narrative: it begins with the combined HV-shots and instance-level best-vs-baseline summary in Fig.~\ref{fig:hv_shots_gap}, then uses summary tables to explain which design choices dominate in each stage, and finally isolates the marginal contribution of the Seed-Selection Methods in Fig.~\ref{fig:marginal_gains}. The full stage-wise heatmaps of method combinations are provided in Appendix~\ref{app:quality_heatmaps} (Fig.~\ref{fig:quality_heatmaps_app}), where the complete adaptive-weight-direction / seed-selection combination grid is shown for reference while the main text focuses on the dominant winners and their mechanism-level interpretation.

\subsection{HV-Shots Curves and Instance-Level Gains}
\label{sec:results:hvshots}

To examine how quality accumulates as the total sampling budget increases, we aggregate the saved trace data into the stage-level HV-shots curves shown in the upper row of Fig.~\ref{fig:hv_shots_gap}. For each dataset, the plot keeps only the single best QEMOO scheme from the rerun benchmark, the corresponding matched random-sampling variant, and the 1-round QAMOO baseline of \citet{kotil2025qamoo}. Specifically, the displayed QEMOO schemes are S6 for DataSet~1: Grid Small, S9 for DataSet~2: Strong Conflict, and S2 for DataSet~3: Grid Large. The vertical axis is $\mathrm{HV}_{\mathrm{best}}-\mathrm{HV}_{t}+1$, where $\mathrm{HV}_{\mathrm{best}}$ is the best observed mean HV within the same dataset and $\mathrm{HV}_{t}$ is the mean HV at cumulative shot count~$t$; the best value is placed at the top, so curves lower in the panel are farther from the stage-best level.

This view clarifies two separate effects. First, the best QEMOO scheme in each dataset stays much closer to the stage-best level than the single-pass baseline, especially on Grid Large where the separation is dramatic. Second, replacing the QAOA sampler in the same scheme with random bitstring sampling causes a large quality drop, showing that the improvement cannot be explained solely by the outer selection logic: the quantum sampling process itself contributes materially to the frontier quality. The lower row of Fig.~\ref{fig:hv_shots_gap} complements this trace-level view by showing the per-instance HV gain of the best QEMOO scheme over the 1-round baseline on all 10 benchmark instances. The remaining subsections explain which method combinations generate these strong traces and instance-level gains in each benchmark stage. Fig.~\ref{fig:marginal_gains} then provides a separate seed-method marginal view, and the appendices explain why the stage-specific winners remain stable under mechanism and hyperparameter analysis.

\subsection{Grid Small ($4 \times 5$ grids)}
\label{sec:results:convex}

On the smaller random-grid benchmark, pushing exploration slightly ahead of historical Pareto anchors remains effective, and the latest rerun benchmark identifies \textbf{S6} (Forward + kNN Sparsity) as the best overall configuration. It improves the mean Hypervolume by $1.32 \times 10^5$ over the baseline, corresponding to a relative gain of about $0.04\%$, and reaches a mean HV of $306.27 \times 10^6$. The top three combinations, S6, S5, and S3, are again very close to one another, which indicates that this moderate-conflict regime is comparatively stable as long as the active adaptive weight direction stays near the original weighted-sum geometry. In this setting, moderate directional extrapolation remains beneficial, while the preferred Seed-Selection Method now slightly favors the kNN-sparsity rule.

\begin{table}[htbp]
\caption{Performance summary of the top three elite-guided configurations for the Grid Small dataset ($4 \times 5$ grid) under the latest rerun raw-HV benchmark. Scheme IDs are defined in Fig.~\ref{fig:overview}. The table lists the mean raw Hypervolume (HV) gain over the 1-round baseline and the win rate (Positive Cases) across all 10 instances. The three best combinations are tightly clustered, with S6 (Forward + kNN Sparsity) giving the largest HV mean gain.}
\label{tab:convex_summary}
\begin{ruledtabular}
\begin{tabular}{lcc}
Scheme & \shortstack{HV Mean \\ Gain} & Positive Cases \\ \hline
S6: Forward + kNN Sparsity & $1.32\times 10^5$ & 10/10 \\
S5: Forward + HV ranking & $1.19\times 10^5$ & 10/10 \\
S3: Inherit + kNN Sparsity & $1.09\times 10^5$ & 10/10 \\
\end{tabular}
\end{ruledtabular}
\end{table}

\subsection{Strong Conflict Benchmark: PBI-inspired Direction Adaptation}
\label{sec:results:pbi}

The Strong Conflict benchmark exposes the limitation of reusing fixed or inherited directions when objectives impose sharply opposing edge preferences. Once the Adaptive Weight-Direction Method is poorly aligned with these trade-offs, extra sampling budget provides only weak benefit. In contrast, the PBI-inspired update improves the search by synthesizing a local geometry-aware effective direction around elite solutions. Operationally, this update is implemented through the mechanism described in Section~\ref{sec:method:line}: the non-linear PBI signal is locally linearized and then mapped back to a weighted-sum-like Ising Hamiltonian for QAOA.
The optimal combination is now \textbf{S9} (PBI + kNN Sparsity), delivering an average hypervolume gain of $7.57 \times 10^6$ over the baseline while remaining positive on all 10 test cases. The top three methods, S9, S8, and S7, all rely on the PBI-inspired adaptive scalarization, which confirms that the decisive improvement comes from the adaptive weight-direction family itself rather than from a secondary difference in seed-selection policy.

The lower row of Fig.~\ref{fig:hv_shots_gap} shows that this best Strong Conflict configuration improves every one of the 10 benchmark instances rather than relying on a small number of outliers. Table~\ref{tab:nonconvex_summary} details the performance of the top combinations and shows that all three leading configurations use the PBI-inspired adaptive scalarization. This pairing of per-instance gains with the top-combination table makes the mechanism clearer: the decisive improvement in this stage comes from the adaptive weight-direction family, while the seed-selection rule mainly refines the final ranking among the PBI-inspired variants.

\begin{table}[htbp]
\caption{Performance summary of the top three configurations for the Strong Conflict dataset under the latest rerun raw-HV benchmark. Scheme IDs are defined in Fig.~\ref{fig:overview}. The metrics include HV mean gain and the number of instances where the method improved upon the baseline out of 10. All three top combinations rely on the PBI-inspired adaptive scalarization, confirming that this geometry-driven direction update is the key ingredient in the Strong Conflict benchmark.}
\label{tab:nonconvex_summary}
\begin{ruledtabular}
\begin{tabular}{lcc}
Scheme & \shortstack{HV Mean \\ Gain} & Positive Cases \\ \hline
S9: PBI + kNN Sparsity & $7.57\times 10^6$ & 10/10 \\
S8: PBI + HV ranking & $7.25\times 10^6$ & 10/10 \\
S7: PBI + BC & $5.44\times 10^6$ & 10/10 \\
\end{tabular}
\end{ruledtabular}
\end{table}

\subsection{Grid Large Evaluation ($6 \times 7$ grids)}
\label{sec:results:mps}

To validate scalability beyond exact state-vector simulation bounds, we deployed the tensor-network backend \texttt{mqmps} on computationally demanding $6 \times 7$ grid topologies. At this elevated complexity, the algorithmic optimum shifts toward conservative reuse of historically good directions rather than aggressive geometric deformation. As shown in Table~\ref{tab:mps_summary}, the \textbf{S2} (Inherit + HV ranking) combination yields the largest HV mean gain, improving on the baseline by $7.64 \times 10^8$ on average, or about $7.50\%$ relatively. The next two methods are S5 (Forward + HV ranking) and S3 (Inherit + kNN Sparsity), which again suggests that the HV-based Seed-Selection Method is the most reliable companion once the problem grows large and the MPS backend introduces approximation noise. This Grid-Large stage is also where the appendix matters most: Appendix~\ref{app:large_mps_hyperparams} revisits the same protocol through controlled scans of the warm-start coefficient, bond dimension, and weight count, so the appendix should be read as an explanation of why this stage winner is stable rather than as a disconnected side study.

\begin{table}[htbp]
\caption{Performance summary of the top three configurations for the Grid Large dataset ($6 \times 7$ grid, \texttt{mqmps} with $\chi=30$) under the latest rerun raw-HV benchmark. Scheme IDs are defined in Fig.~\ref{fig:overview}. Evaluated metrics include HV mean gain and positive case count. At this scale, the conservative S2 (Inherit + HV ranking) combination remains the strongest overall strategy. Appendix~\ref{app:large_mps_hyperparams} then explains how this regime behaves under the key Grid-Large hyperparameters.}
\label{tab:mps_summary}
\begin{ruledtabular}
\begin{tabular}{lcc}
Scheme & \shortstack{HV Mean \\ Gain} & Positive Cases \\ \hline
S2: Inherit + HV ranking & $7.64\times 10^8$ & 10/10 \\
S5: Forward + HV ranking & $7.19\times 10^8$ & 10/10 \\
S3: Inherit + kNN Sparsity & $6.68\times 10^8$ & 10/10 \\
\end{tabular}
\end{ruledtabular}
\end{table}

\begin{figure*}[htbp]
\centering
\includegraphics[width=\textwidth]{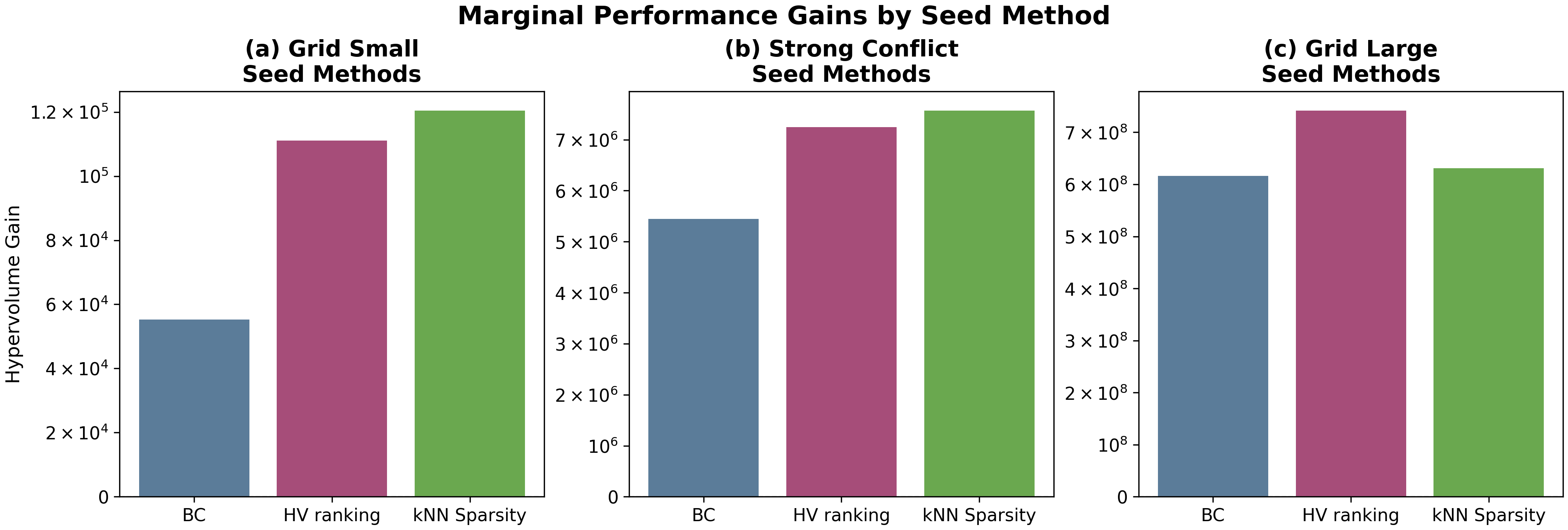}
\caption{Marginal Hypervolume gains of the Seed-Selection Methods in the rerun raw-HV benchmark. The three panels correspond to the Grid Small, Strong Conflict, and Grid Large benchmark stages. Each bar reports the baseline-relative HV gain averaged over the stage-appropriate Adaptive Weight-Direction Methods: Inherit and Forward for Grid Small and Grid Large, and PBI-inspired direction adaptation for Strong Conflict. This figure therefore isolates the average contribution of the Seed-Selection Method (BC, HV ranking, or kNN Sparsity), where BC denotes Balanced Coverage. It is complementary to the lower row of Fig.~\ref{fig:hv_shots_gap}: Fig.~\ref{fig:hv_shots_gap} reports per-instance gains for the single best scheme in each dataset, whereas this figure collapses across stage-appropriate direction choices to compare seed-selection rules.}
\label{fig:marginal_gains}
\end{figure*}

Fig.~\ref{fig:marginal_gains} adds a seed-selection marginal view to the stage-wise winner tables. In Grid Small, kNN Sparsity and HV ranking both outperform BC on average, with kNN Sparsity giving the largest marginal gain, consistent with the S6 winner in Table~\ref{tab:convex_summary}. In the Strong Conflict stage, all bars are computed within the PBI-inspired row, so the plot should not be read as evidence that seed selection alone explains the stage winner; rather, it shows that kNN Sparsity provides the strongest refinement once the geometry-aware direction update is active. In Grid Large, HV ranking has the largest seed-method marginal gain, matching the dominance of S2 and S5 in Table~\ref{tab:mps_summary}. Thus Fig.~\ref{fig:marginal_gains} separates the seed-selection effect from the per-instance best-scheme gains shown in Fig.~\ref{fig:hv_shots_gap}.

\subsection{Robustness and Absolute Superiority}
\label{sec:results:robustness}

To substantiate the superiority of the best multi-round configurations, we examined the best-performing configuration within each dataset and the per-instance winner patterns embedded in the rerun benchmark outputs. The pattern is sharp: in the Strong Conflict stage, every winning configuration relies on the PBI-inspired adaptive update, whereas on Grid Large the majority of wins concentrate on S2 (Inherit + HV ranking). The lower row of Fig.~\ref{fig:hv_shots_gap} zooms in on this claim at the instance level by pairing, for each dataset separately, the single best multi-round configuration from the rerun benchmark with the corresponding 1-round baseline on all 10 test cases. The plot therefore does not average the 10 cases into a single point; instead, each bar is one concrete instance-level HV gain over the baseline. From the reader's perspective, this closes the main-text story: Fig.~\ref{fig:hv_shots_gap} shows both the dynamic budget advantage and the concrete per-instance improvements, Tables~\ref{tab:convex_summary}--\ref{tab:mps_summary} identify the stage winners, and Fig.~\ref{fig:marginal_gains} explains how the seed-selection component contributes on average within each stage. The appendices then unpack the mechanism and parameter choices behind this pattern.

\section{Discussion}
\label{sec:discussion}

The results suggest that the advantage of QEMOO comes from a two-level information-reuse mechanism. At the parameter level, transferred QAOA angles provide a useful cross-instance prior: nearby scalarized subproblems often share correlated variational parameters, reducing the optimization overhead associated with solving many weight directions. At the state level, the multi-round warm-start loop converts previously discovered non-dominated elites into sampling priors for later rounds. The HV-shots curves are consistent with this interpretation: when the same adaptive weight-direction and seed-selection logic is kept fixed, QAOA sampling remains separated from the matched random controls, indicating that the quantum sampler contributes beyond the outer classical feedback loop alone.

The stage-dependent winners also clarify the role of the modular design. QEMOO should not be interpreted as a single universal configuration, but as a framework whose most effective instantiation depends on the geometry of the Pareto front and the available sampling budget. In the Strong Conflict stage, fixed directions can become poorly aligned with local frontier geometry, making the PBI-inspired adaptive update particularly useful. By contrast, in the Grid Small and Grid Large stages, inherited or forward-updated directions remain competitive, suggesting that conservative reuse of previously effective directions can be preferable when the frontier geometry is smoother or when tensor-network approximation noise becomes relevant. Importantly, this distinction concerns how effective directions are generated; after local linearization, the PBI-inspired update still enters QAOA through an effective weighted-sum Ising Hamiltonian.

Several limitations should be kept in mind. First, the present study is based on simulated QAOA sampling and MPS-based large-scale evaluation, so hardware noise, readout errors, connectivity constraints, and real device queue-time costs remain to be tested. Second, hypervolume is used as the primary quality indicator, and its absolute value depends on objective scaling and reference-point choices; for this reason, we emphasize matched-budget comparisons, random-sampling controls, and per-instance no-regression tests. Third, the Grid-Large hyperparameter scans show that settings such as $\chi=30$, $c=0.4$, and larger weight counts are the strongest tested choices in our benchmark suite, not universal optima. More broadly, the all-9 weight-pool study indicates that weight-pool geometry affects both final quality and round-by-round recovery, but does not collapse to a simple monotone rule.

These observations point to a broader design principle for quantum multi-objective optimization: useful quantum samplers should be embedded in feedback-driven, geometry-aware, and budget-aware optimization loops rather than used only as single-pass solvers for fixed scalarizations. Future work can build on this principle by learning adaptive direction and warm-start policies from previous runs, extending the framework to deeper and hardware-calibrated ansatzes, and testing the full protocol under realistic noise and sampling constraints.

\section{Conclusion}
\label{sec:conclusion}

In this work, we introduced QEMOO, a modular multi-round framework organized around two complementary method dimensions: Adaptive Weight-Direction Methods and Seed-Selection Methods. The framework combines transferred QAOA angles across problem instances with warm-start feedback across rounds, incorporates a PBI-inspired adaptive direction update for strongly conflicting objective structures, and is supported by HV-shots evidence showing that the quantum sampler contributes beyond what can be reproduced by matched random sampling under the same budget.

The central empirical lesson is not that a single universal recipe dominates every benchmark stage, but that effective multi-objective quantum optimization must match its algorithmic bias to the Pareto geometry and available sampling budget. This is where the modular structure of QEMOO is most useful: by separating adaptive direction construction from elite-solution transfer, it turns stage-dependent behavior into a diagnosable design space rather than a fixed heuristic. Future work can build on this design space in two directions: learning weight-direction and warm-start policies from previous runs, and testing deeper, hardware-calibrated ansatzes under realistic device and sampling constraints.

\section*{Code and Data Availability}

The code and data needed to reproduce the numerical results in this work are available at \url{https://github.com/luoml5/qemoo}.

\begin{acknowledgments}
This work is supported by Science, Technology and Innovation Bureau of Shenzhen Municipality.
\end{acknowledgments}

\bibliographystyle{apsrev4-2}
\nocite{apsrev42Control}
\bibliography{ref}

\appendix
\onecolumngrid

\section{Full Heatmaps of Method Combinations}
\label{app:quality_heatmaps}

For completeness, we provide the full comparison of stage-wise method combinations for the rerun raw-HV benchmark. This appendix is the detailed counterpart of the main-text narrative: the main text highlights the dominant winners, while this section exposes the complete adaptive-weight-direction / seed-selection combination grid behind those patterns.

\begin{figure}[htbp]
    \centering
    \includegraphics[width=0.4\columnwidth]{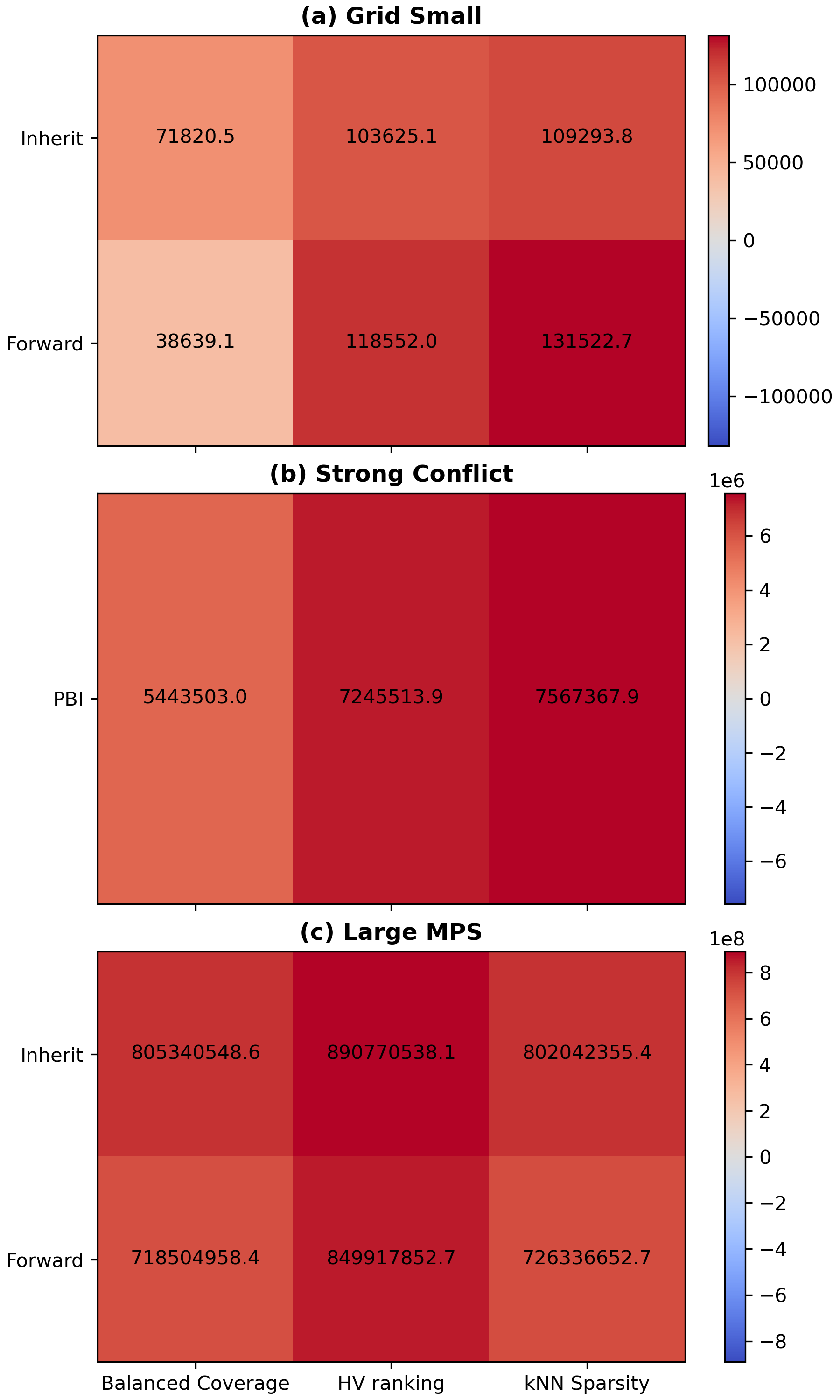}
    \caption{Performance heatmaps detailing the hypervolume improvements across the rerun raw-HV benchmark suite: (a) Grid Small, (b) Strong Conflict, and (c) Grid Large. In each subplot, the horizontal axis delineates the Seed-Selection Methods (BC, HV ranking, kNN Sparsity), while the vertical axis lists the stage-appropriate Adaptive Weight-Direction Methods retained in the latest benchmark: Inherit and Forward for Grid Small and Grid Large, and PBI for the strong-conflict stage. Cell colors represent the mean raw Hypervolume (HV) gain relative to the 1-round baseline, with warm colors indicating positive gains and cool colors denoting regressions. The heatmaps show that the preferred direction/seed combination is strongly stage dependent.}
    \label{fig:quality_heatmaps_app}
\end{figure}

These heatmaps should be read as the complete background grid behind the stage-wise story in the main text, rather than as an alternative ranking device. Their main value is to show which combinations remain near-optimal within each stage and which combinations degrade systematically once the adaptive weight-direction method is mismatched to the underlying geometry. In this sense, Appendix~\ref{app:quality_heatmaps} complements the winner tables and Fig.~\ref{fig:hv_shots_gap}'s best-scheme panels by making the full comparison space explicit.

\section{Impact of Weight Direction Distributions}
\label{app:weights}

This section provides the mechanism-level support for the main benchmark by showing how the geometry of the weight pool affects warm-start transfer. Here we fix the Grid-Large setting to \texttt{inherit + hv}, \texttt{mqmps} with $\chi=30$, and $c=0.4$, and vary only the weight-pool distribution $\Lambda$. The all-9 comparison therefore complements the main-text benchmark: the main text keeps \texttt{dirichlet\_uniform} as the default pool, while this appendix asks how alternative pool geometries change the balance between absolute final quality and warm-start gain.

\begin{itemize}
\item \textbf{Dirichlet family:} 
  Samples drawn from $\text{Dir}(\alpha, \ldots, \alpha)$ with $\alpha \in \{0.2, 1.0, 5.0\}$, controlling the concentration: $\alpha < 1$ yields corner-heavy distributions, $\alpha = 1$ gives the uniform distribution on the simplex, and $\alpha > 1$ concentrates mass near the centroid. A \emph{mixed} variant combines equal proportions of all three.
\item \textbf{Normalized continuous distributions:} 
  \emph{Uniform box} (components drawn i.i.d. from $U(0,1)$ then normalized), \emph{Gaussian clip} (truncated Gaussian $\mathcal{N}(0.5, 0.2)$), and \emph{log-normal} entries.
\item \textbf{Structured designs:} 
  \emph{Simplex lattice} with deterministic grid points plus random fill, and \emph{edge--corner mix} with handcrafted boundary templates.
\end{itemize}

\begin{figure}[htbp]
\centering
\includegraphics[width=0.9\columnwidth]{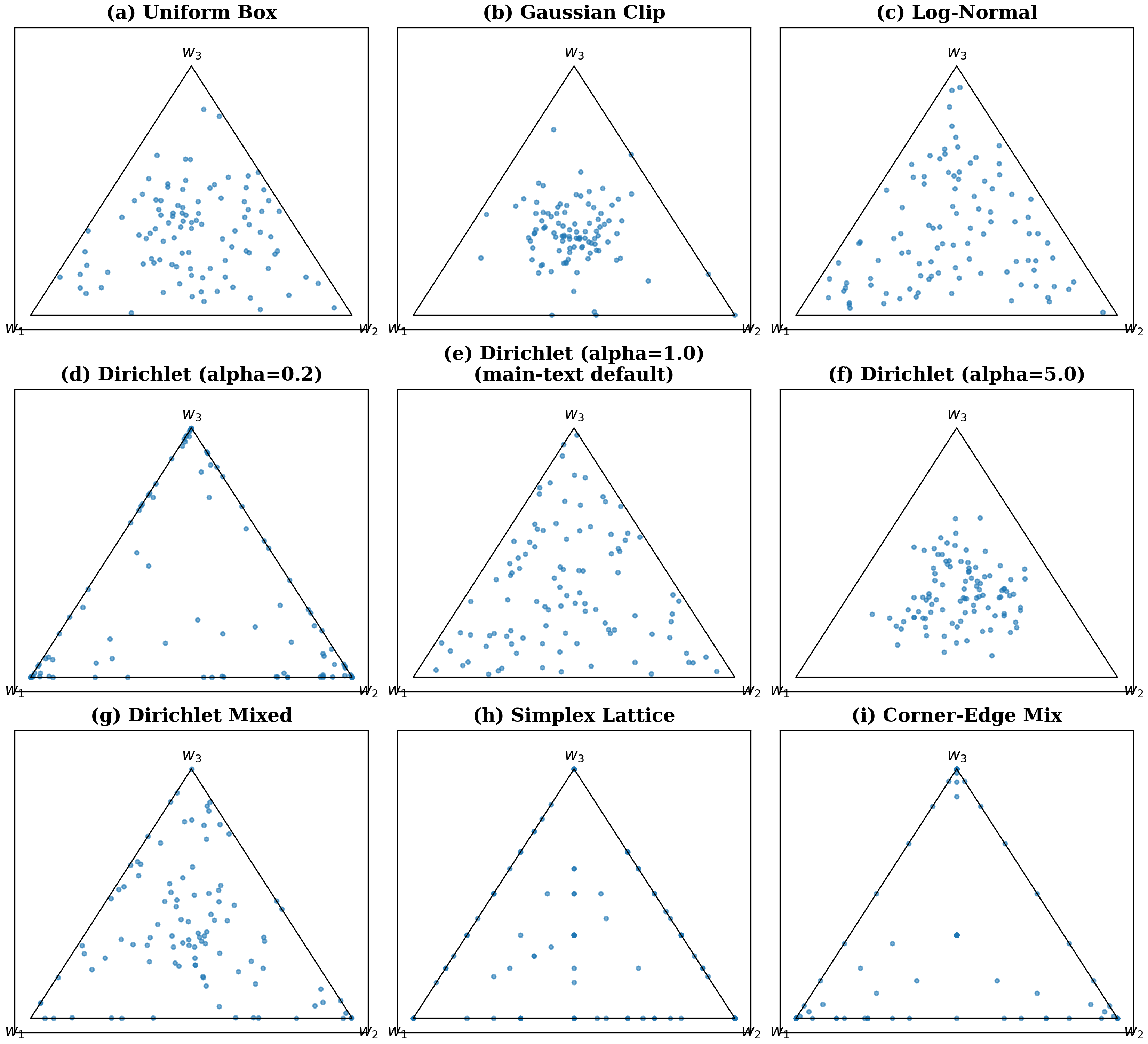}
\caption{Simplex-style visualization of the nine weight-pool schemes used in the latest all-9 Grid-Large analysis. Each panel is generated from the corresponding real pool file in the rerun suite, with vertex labels $w_1$, $w_2$, and $w_3$ marking single-objective extremes. The figure covers normalized continuous pools (\texttt{uniform\_box}, \texttt{gaussian\_clip}, \texttt{lognormal}), the Dirichlet family (\texttt{dirichlet\_sparse}, \texttt{dirichlet\_uniform}, \texttt{dirichlet\_center}, \texttt{dirichlet\_mixed}), and structured designs (\texttt{simplex\_lattice}, \texttt{edge\_pair\_corner\_mix}). The \texttt{dirichlet\_uniform} panel is flagged as the main-text default pool.}
\label{fig:pool_simplex}
\end{figure}

To quantify the geometric attributes of a weight pool, we track three distribution-level statistics: (i)~\emph{Normalized entropy} $\bar{H}$ measuring distribution uniformity, (ii)~\emph{Mean maximum component} $\bar{\lambda}_{\max}$ tracking extremity bias, and (iii)~\emph{Mean pairwise $L_2$ distance} $\bar{d}$. Together these statistics summarize whether a pool is smooth and center-oriented, sharply corner-biased, or more structured along edges and lattice directions. Figure~\ref{fig:pool_simplex} visualizes the nine schemes using the same simplex-style view as the rest of the paper; the \texttt{dirichlet\_uniform} panel is marked as the main-text default.

The latest all-9 Grid-Large rerun reveals a clearer trade-off than the earlier uniform-family study. By joint final quality, the strongest pool is \texttt{dirichlet\_sparse}, with mean baseline HV $1.0603$, mean warm HV $1.1175$, and mean gain $0.0572$. By pure warm-start gain, however, the strongest pool is \texttt{gaussian\_clip}, with mean gain $0.0883$ even though its final joint score is lower. The default \texttt{dirichlet\_uniform} pool remains competitive and stable, but in this focused Grid-Large comparison it is no longer the best performer by either joint score or gain alone. This is the main qualitative update: pool geometry matters, and the pool that maximizes absolute final HV need not be the same one that maximizes warm-start amplification.

\begin{figure}[htbp]
\centering
\includegraphics[width=0.5\columnwidth]{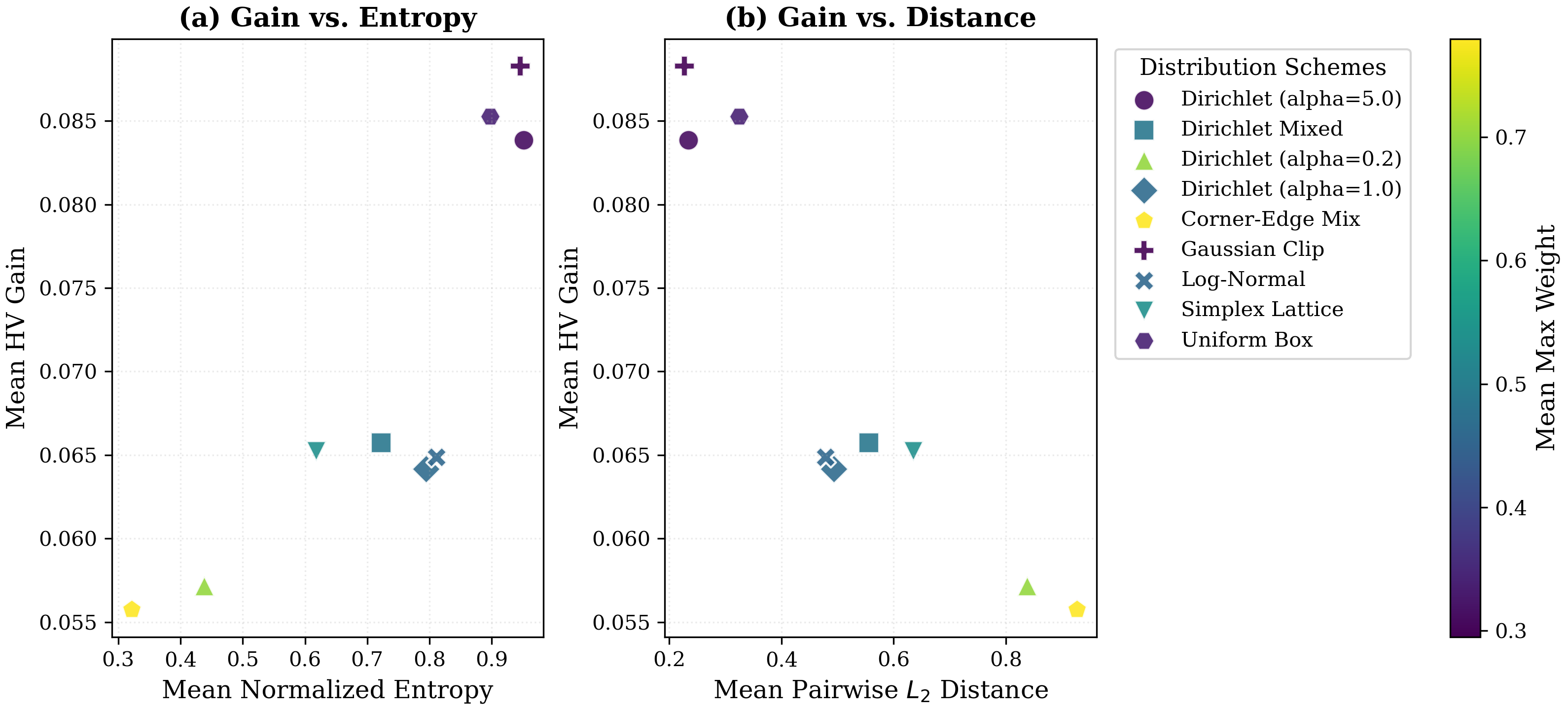}
\caption{Relationship between weight-pool geometry and warm-start gain in the latest all-9 Grid-Large rerun. Panel~(a) plots mean HV gain against normalized entropy, and Panel~(b) plots mean HV gain against mean pairwise $L_2$ distance. Marker shape identifies the pool scheme, and marker color represents the mean maximum weight ($\bar{\lambda}_{\max}$). The two panels show that smoother pools often support more consistent transfer, but the strongest final absolute quality and the strongest gain can still arise from different pool geometries.}
\label{fig:gain_smoothness}
\end{figure}

\begin{figure}[htbp]
\centering
\includegraphics[width=0.5\columnwidth]{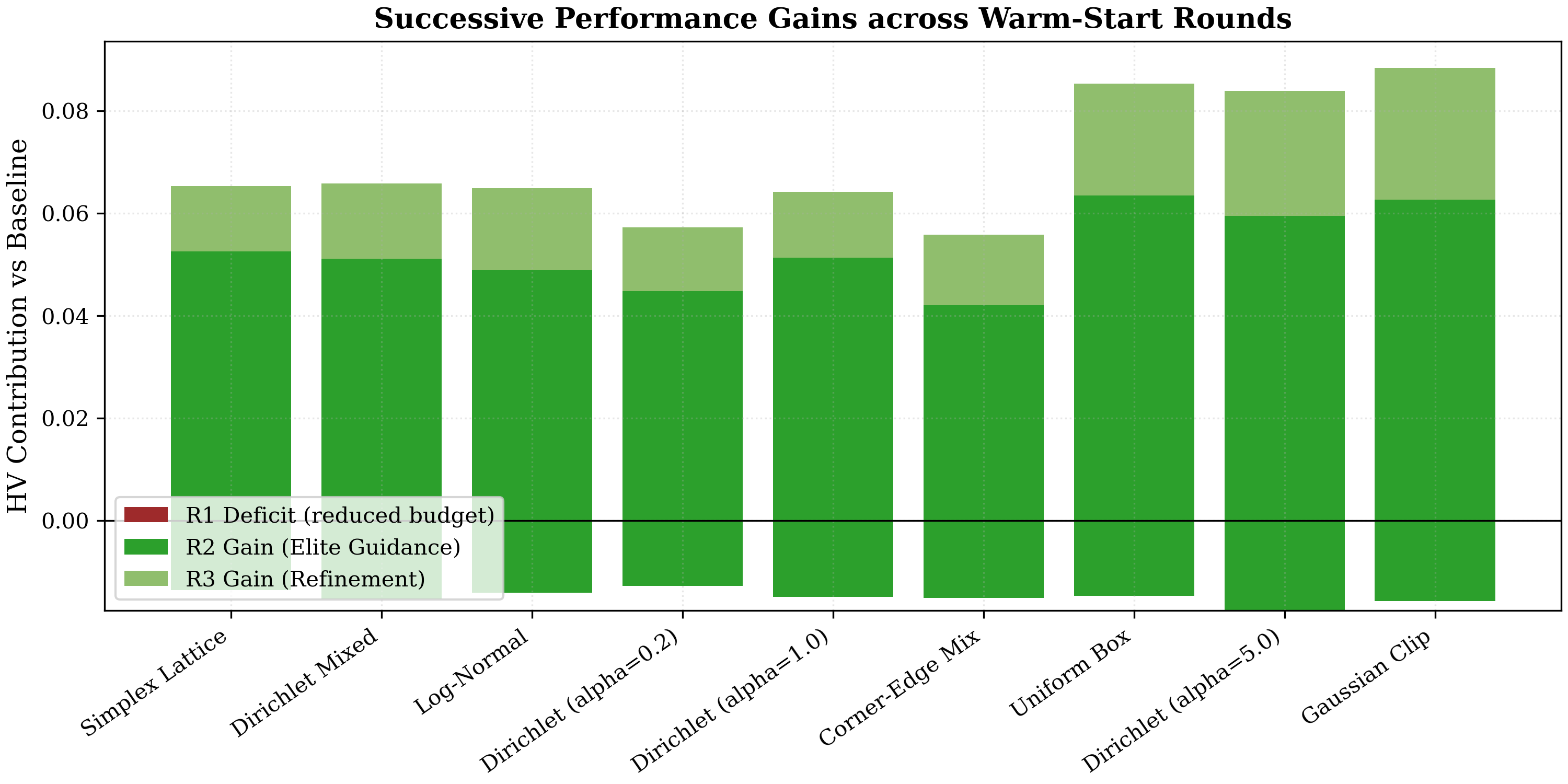}
\caption{Round-wise decomposition of warm-start gains for the nine weight-pool schemes in the latest all-9 Grid-Large rerun. Each bar is expressed relative to the matched baseline: the red segment is the Round-1 deficit caused by splitting the warm budget across rounds, while the two green segments show the recoveries contributed by Round~2 and Round~3. The dominant pattern is ``deficit--recovery'': the later warm rounds more than compensate for the initial loss, but the size of that recovery remains pool dependent.}
\label{fig:round_recovery}
\end{figure}

Figure~\ref{fig:gain_smoothness} should therefore be read as a two-view mechanism plot rather than as a one-number rule such as ``smoother is always better.'' Panel~(a) relates gain to normalized entropy, while Panel~(b) relates gain to mean pairwise distance. Together they show that smoother pools can improve transfer consistency by placing nearby directions closer in the simplex, but the latest all-9 rerun also shows that edge- and corner-emphasizing pools can remain highly competitive in final quality. Likewise, Fig.~\ref{fig:round_recovery} is best read as a round-wise decomposition of where the warm-start advantage is accumulated, and the updated ranking highlights a split between absolute-quality winners and gain-oriented winners.

The round-wise bar decomposition in Fig.~\ref{fig:round_recovery} can also be summarized numerically by tracking the mean cumulative HV after each round together with the round-to-round increments. Table~\ref{tab:round_hv_change} is therefore best read as a mechanism summary: it shows how different pool geometries redistribute improvement between the initial deficit of Round~1 and the subsequent recoveries in Round~2 and Round~3. Combined with the latest all-9 ranking, the broader lesson is not that one geometric family dominates every criterion, but that pool design changes both the final score and the temporal pattern of recovery.

\begin{table}[htbp]
\centering
\caption{Round-wise cumulative HV statistics for the nine weight-pool schemes. ``Round 1'' is the mean cumulative HV after the first pass, while ``Round 2 gain'' and ``Round 3 gain'' report the mean incremental improvement over the previous round. The table highlights how the warm-start feedback loop redistributes the fixed budget across rounds.}
\label{tab:round_hv_change}
\small
\begin{tabular}{lrrrrrr}
\hline
Scheme & Baseline & Round 1 & Round 2 & Round 3 & Round 2 gain & Round 3 gain \\
\hline
simplex\_lattice & 1.0534 & 1.0398 & 1.1058 & 1.1186 & 0.0661 & 0.0128 \\
dirichlet\_uniform & 1.0531 & 1.0381 & 1.1044 & 1.1173 & 0.0663 & 0.0128 \\
dirichlet\_mixed & 1.0528 & 1.0375 & 1.1039 & 1.1185 & 0.0664 & 0.0146 \\
dirichlet\_center & 1.0125 & 0.9949 & 1.0719 & 1.0963 & 0.0770 & 0.0244 \\
gaussian\_clip & 0.9947 & 0.9790 & 1.0573 & 1.0830 & 0.0783 & 0.0256 \\
uniform\_box & 1.0152 & 1.0005 & 1.0787 & 1.1005 & 0.0782 & 0.0218 \\
lognormal & 1.0533 & 1.0392 & 1.1022 & 1.1182 & 0.0630 & 0.0160 \\
dirichlet\_sparse & 1.0603 & 1.0475 & 1.1051 & 1.1175 & 0.0576 & 0.0124 \\
edge\_pair\_corner\_mix & 1.0592 & 1.0440 & 1.1012 & 1.1149 & 0.0571 & 0.0137 \\
\hline
\end{tabular}
\end{table}

\section{Large-MPS Hyperparameter Sensitivity}
\label{app:large_mps_hyperparams}

The Grid-Large experiments are the most parameter-sensitive part of the rerun benchmark, so we group the key hyperparameter studies together here. This appendix is the parameter-level support for the Grid Large winner in the main text: we first identify the best warm-start mixing coefficient $c$, then inspect the bond-dimension scaling through $\chi$, and finally study how the number of sampled weights affects the best configuration.

\subsection{Selection Strategy for Mixing Coefficient $c$}
\label{app:warmc}

\begin{figure}[htbp]
\centering
\includegraphics[width=0.5\columnwidth]{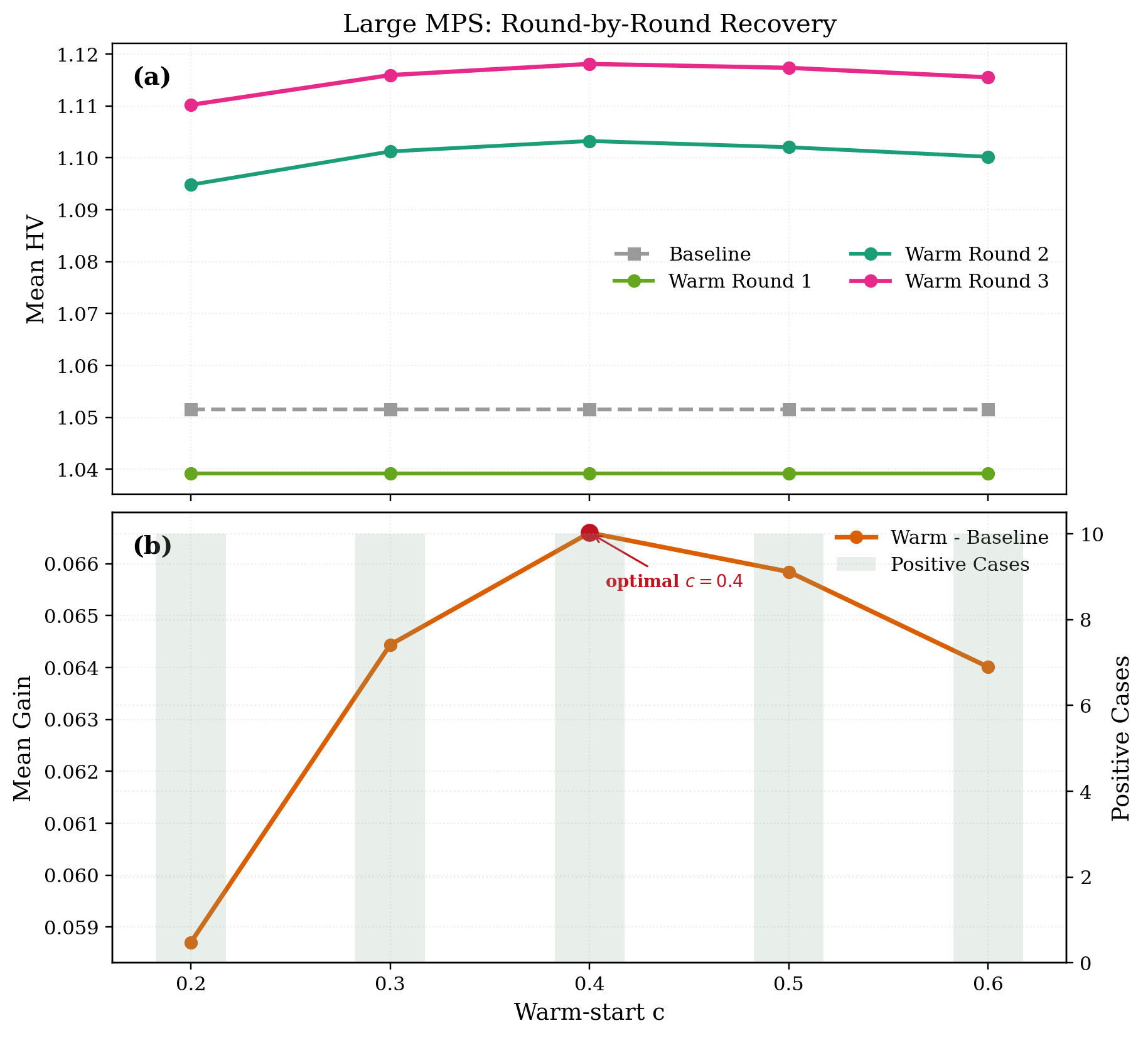}
\caption{Mean HV vs.\ warm-start coefficient $c$ with round-recovery and gain/positive-case decomposition. The baseline (dashed) is independent of $c$, and the warm-start optimum occurs at $c = 0.4$.}
\label{fig:warmc_overview}
\end{figure}

The warm-start bias coefficient $c \in [0, 1]$ serves as the primary control for the exploration--exploitation trade-off. We explore its impact by scanning from $c=0.1$ (weak bias) to $c=0.8$ (strong exploitation) across all test instances.
As shown in Figure~\ref{fig:warmc_overview}, the resulting Hypervolume follows a clear unimodal distribution peaking at $c \approx 0.4$. Below this threshold, the quantum search remains too close to a cold start, failing to fully utilize the guidance of elite seeds. Above $c=0.6$, performance declines as the bias becomes over-restrictive.

The upper panel in Figure~\ref{fig:warmc_overview} shows the round-recovery trajectories across $c$, while the lower panel decomposes the scan into mean HV gain and positive-case count. Round 1 is identical for all values because warm-starting only begins in later rounds, so the performance differences arise entirely from the feedback mechanism. The gain curve peaks at $c=0.4$, and the positive-case bars indicate that this setting produces the most consistently beneficial warm-start behavior across instances. Our results identify $c=0.4$ as a robust "sweet spot" that biases the search into high-quality regions of the Ising landscape without sacrificing the diversity of the final Pareto set.

\subsection{Bond-Dimension Scaling for the Best Large-MPS Case}
\label{app:chi}

\begin{figure}[htbp]
\centering
\includegraphics[width=\columnwidth]{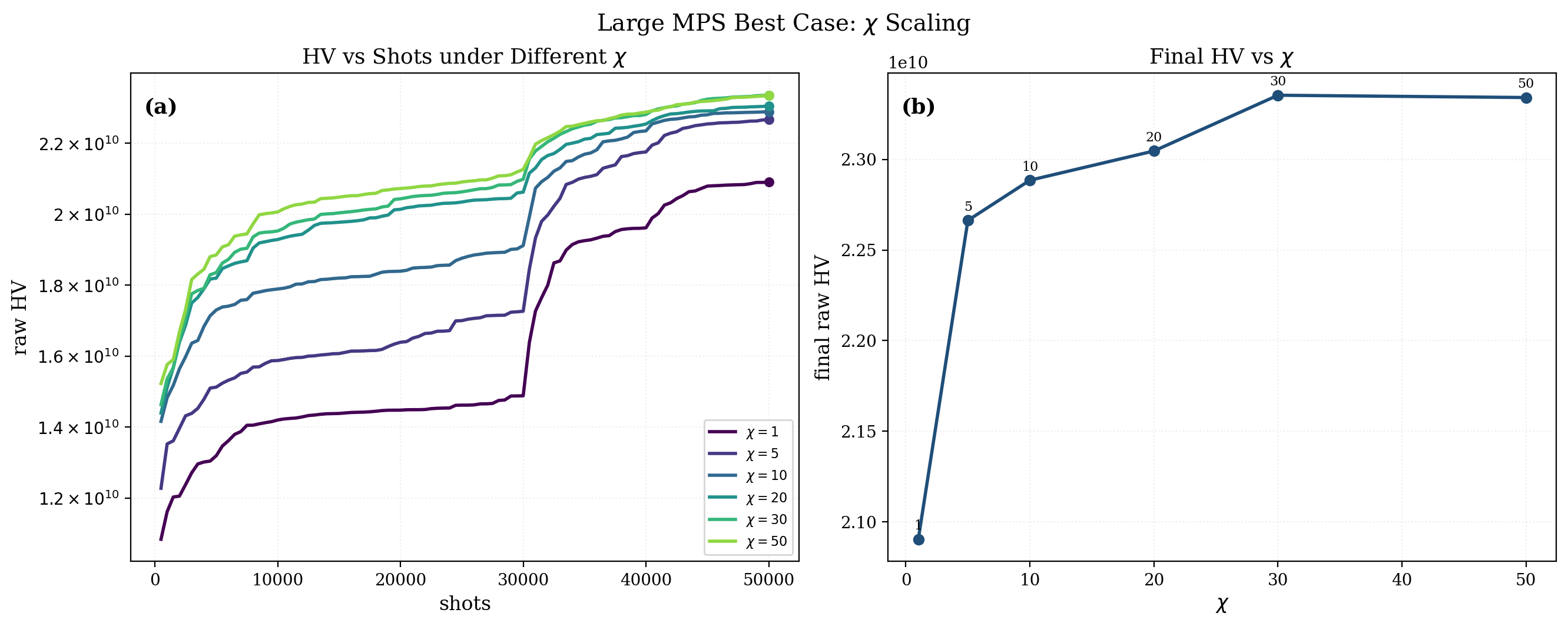}
\caption{Bond-dimension scaling for the best Grid-Large case. The left panel shows the HV-vs-shots trajectories for different bond dimensions $\chi$, and the right panel reports the final mean HV as a function of $\chi$. The rerun indicates that the best final value is achieved near $\chi=30$.}
\label{fig:chi_scaling}
\end{figure}

To assess how tensor-network approximation quality interacts with the best Grid-Large configuration, we vary the bond dimension $\chi$ while keeping the same warm-start protocol and best observed direction policy.
The left panel in Fig.~\ref{fig:chi_scaling} shows the HV-vs-shots trajectories for different $\chi$ values, and the right panel summarizes the final HV as $\chi$ changes.
The key takeaway is that the Grid-Large result is already stable at moderate bond dimension, with the highest final mean HV appearing near the latest $\chi=30$ rerun setting.

\subsection{Weight-Count Scaling for the Best Large-MPS Case}
\label{app:numweights}

\begin{figure}[htbp]
\centering
\includegraphics[width=\columnwidth]{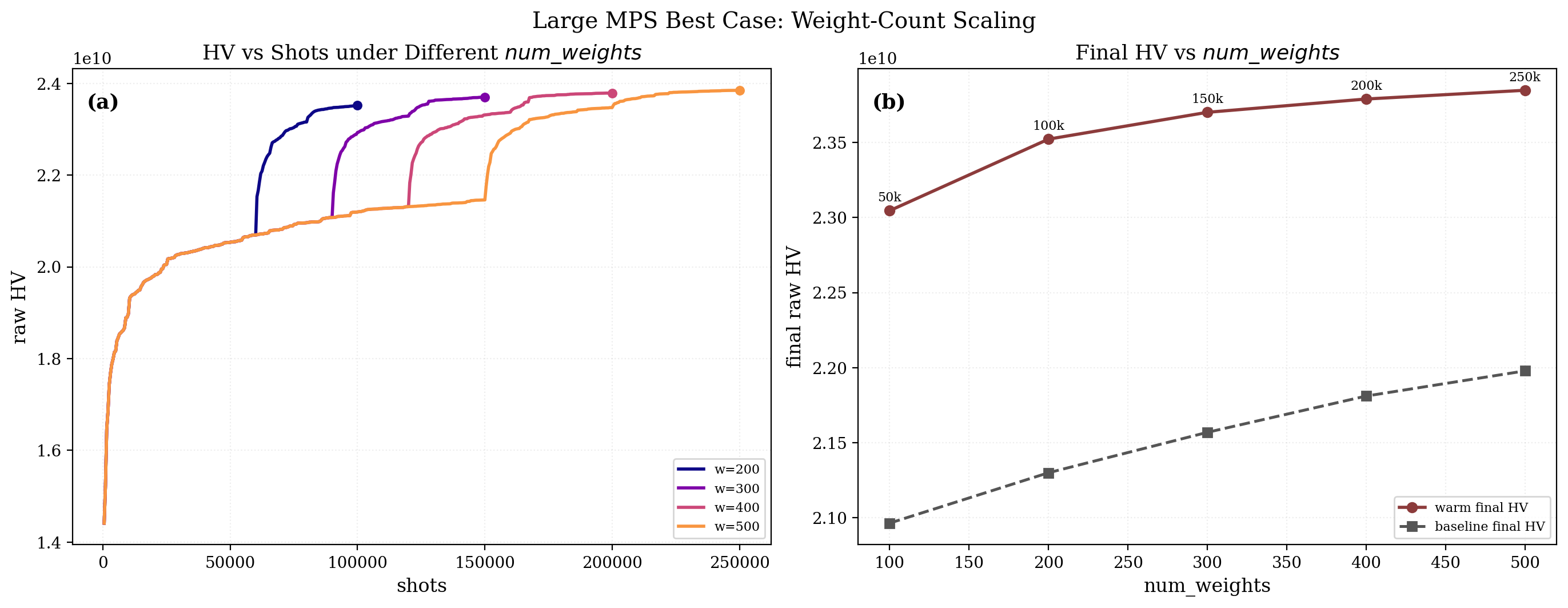}
\caption{Weight-count scaling for the best Grid-Large case. The left panel shows the HV-vs-shots trajectories for different numbers of sampled weights, and the right panel compares the final HV against the matched baseline. The right panel omits the auxiliary warm-minus-baseline curve to keep the comparison focused.}
\label{fig:numweights_scaling}
\end{figure}

To complement the warm-start coefficient analysis, we also examine how the best Grid-Large configuration changes as the number of sampled weights varies while keeping the same transferred parameters and sampling protocol.
The left panel in Fig.~\ref{fig:numweights_scaling} shows the HV-vs-shots trajectories for different weight counts, and the right panel compares the final HV achieved by each setting against its matched baseline.
This view makes the scaling trend easier to interpret: more weights provide broader directional coverage, but the improvement saturates as the sample budget grows.

\section{Impact of Shot Counts}
\label{app:budget}

\begin{figure}[htbp]
\centering
\includegraphics[width=0.5\columnwidth]{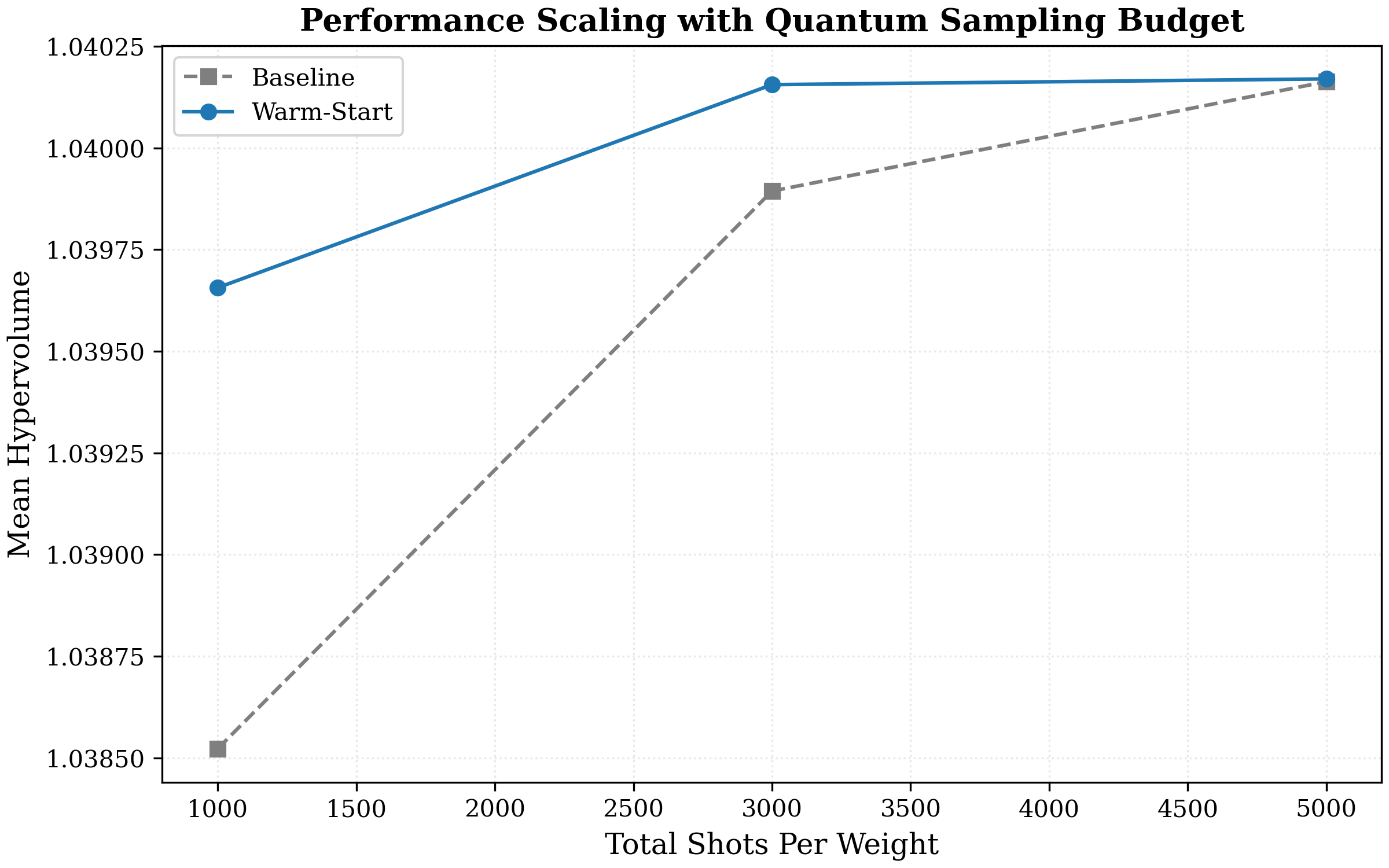}
\caption{Hypervolume scaling behavior as a function of the per-weight sampling budget $s_{\text{total}}$. The x-axis identifies the total shots per direction $(1000, 3000, 5000)$. At high budgets, the baseline and warm-start methods natively converge as the exploration space is exhaustively sampled, proving that our framework's primary advantage lies in shot-restricted hardware environments.}
\label{fig:budget_hv}
\end{figure}

\begin{figure}[htbp]
\centering
\includegraphics[width=0.5\columnwidth]{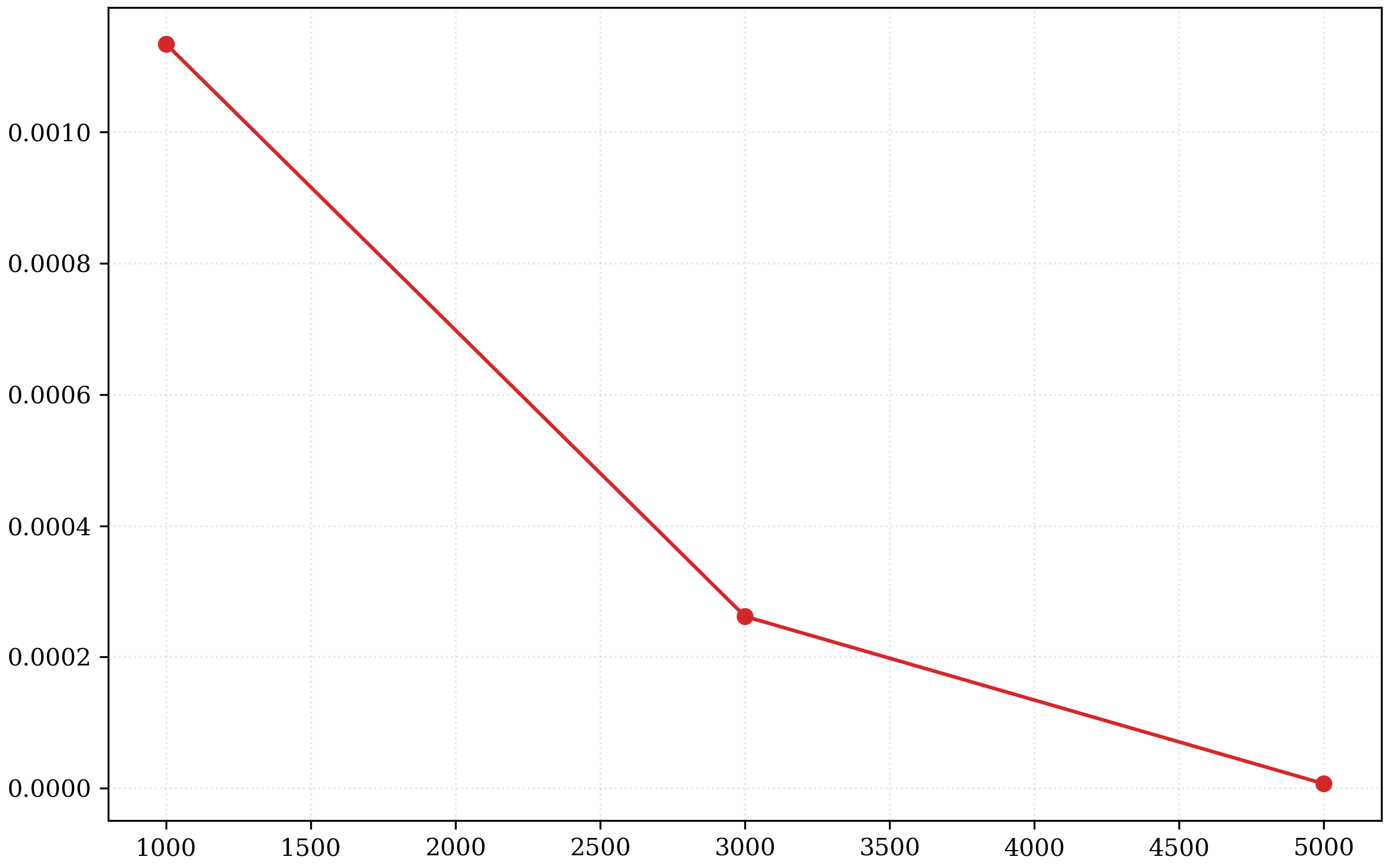}
\caption{Analysis of mean $\Delta$HV gain (left y-axis, red line) and positive instance counts (right y-axis, transparent bars) across varying budgets. The data confirms that the elite-guided warm-start method is most advantageous at moderate budgets, providing consistent improvement across all 10 instances while maintaining high absolute gains.}
\label{fig:budget_delta}
\end{figure}

This section moves from mechanism and parameter support to resource robustness. To determine the efficiency of warm-starting as a function of total computational resources, we explore three budget levels: $s_{\text{total}} \in \{1000, 3000, 5000\}$ shots per direction.
The scaling results in Figure~\ref{fig:budget_hv} and Figure~\ref{fig:budget_delta} demonstrate that the warm-start advantage ($\Delta\hvol$) is most pronounced in the resource-constrained regime ($s_{\text{total}} = 1000$). At this level, elite-guided rounds provide a consistent and substantial improvement ($\Delta\hvol \approx 0.0011$) across $10/10$ test cases.

As the budget increases towards 5000 shots, the gain diminishes significantly. This behavior implies that the baseline single-pass QAOA eventually explores the low-energy modes of the Ising landscape exhaustively given enough shots, leaving fewer marginal improvements for the feedback loop to capture. However, in the context of NISQ-era hardware where quantum sampling is inherently expensive and restricted by decoherence or queue times, our framework provides a critical mechanism for achieving high-fidelity Pareto fronts at much lower shot counts than brute-force sampling would require.

\section{Quantum vs. Random Sampling}
\label{app:qvr}

This appendix complements the main-text HV-shots analysis by tabulating the matched-budget quantum-versus-random comparison directly. We compare all QAOA variants against classical uniform random sampling and reuse the same matched-budget data source as the updated HV-shots analysis in Fig.~\ref{fig:hv_shots_gap}, ensuring that Appendix~\ref{app:qvr} is numerically consistent with the latest rerun benchmark rather than with an older summary table. The table below reports the same six directional configurations used in the current quantum-vs-random study.

The results show a categorical separation between quantum and classical methods: all QAOA configurations achieve a mean Hypervolume of approximately $1.040$, whereas random sampling struggles to exceed $0.980$. This significant performance gap confirms that the high-quality Pareto fronts reported throughout this study are fundamentally driven by the quantum algorithm's ability to sample low-energy Ising states, rather than a mere consequence of the weighted-sum decomposition. Furthermore, within the quantum methods, the 3-round warm-start configuration discovers the highest count of unique non-dominated solutions, which means that the feedback loop actively broadens the discovery of diverse trade-off points on the Pareto manifold.

\begin{table}[htbp]
\caption{Comparative performance of Quantum (QAOA) vs. Classical (Random) sampling under the same matched-budget setup used in the updated HV-shots analysis in Fig.~\ref{fig:hv_shots_gap}. The table lists the mean Hypervolume achieved and the number of unique non-dominated solutions discovered (ND). The quantum variants remain decisively better than the random baselines under the same budget-parity comparison.}
\label{tab:qvr}
\begin{ruledtabular}
\begin{tabular}{lccc}
Method & Mean HV & Mean ND \\ \hline
q\_single ($n_w{=}400$) & 1.039933 & 35099 \\
q\_single ($n_w{=}200$) & 1.039541 & 31849 \\
q\_dir ($n_w{=}200$, 3-R) & 1.039429 & 40814 \\
r\_dir ($n_w{=}200$, 3-R) & 0.981069 & 1888 \\
r\_single ($n_w{=}400$) & 0.979842 & 1869 \\
r\_single ($n_w{=}200$) & 0.975752 & 1925 \\
\end{tabular}
\end{ruledtabular}
\end{table}

\section{Derivation of the PBI-Inspired Adaptive Scalarization}
\label{app:pbi_derivation}

The Penalty-based Boundary Intersection (PBI) method evaluates a solution vector $\bm{f}(\bm{x})$ relative to an ideal reference point $\bm{z}^*$. Assuming a weight vector normalized to unit length $\bm{u} = \bm{\lambda}/\|\bm{\lambda}\|$, the PBI scalarization is defined as
\begin{equation}
g^{\text{PBI}}(\bm{x}) = d_1 + \theta_p d_2,
\end{equation}
where $d_1 = \bm{u}^\top (\bm{f}(\bm{x}) - \bm{z}^*)$ explicitly captures convergence along the target direction, and $d_2 = \|\bm{f}(\bm{x}) - \bm{z}^* - d_1 \bm{u}\| \equiv \|\bm{d}_2\|$ measures orthogonal diversity penalty.

Mapping this non-linear evaluation into the quantum cost unitary $U_C = e^{-i \gamma H_C}$ is theoretically challenging. While standard combinatorial components intrinsically form quadratic Ising Hamiltonians ($Z_i Z_j$), constructing a target operator for $d_2^2$ (to bypass the square root) inevitably involves expanding $(\bm{f}(\bm{x}) - \bm{z}^* - d_1 \bm{u})^2$. The multiplication of separate two-body objective Hamiltonians $f_i(\bm{x}) f_j(\bm{x})$ natively spawns massive four-body interaction terms ($Z_i Z_j Z_k Z_l$). These higher-order interactions are fundamentally incompatible with near-term quantum processors without resorting to highly resource-intensive ancilla-based quadratization protocols.

To circumvent this hardware penalty and preserve the strict two-body $H_C$ requirement, our warm-start framework deploys a dynamic, first-order linear approximation (Expectation-based Linearization) derived directly from the classical feedback loop. During multi-round sampling, each algorithmic round defines an elite reference seed $\bm{x}_{\text{ref}}$ with known classical multi-objective evaluations $\bm{f}_{\text{raw}} = \bm{f}(\bm{x}_{\text{ref}})$. 

Rather than hardcoding the geometric $d_2^2$ penalty into the global unitary, we locally linearize the PBI function using its exact geometric gradient at $\bm{f}_{\text{raw}}$. Let $\bm{d} = \bm{f}_{\text{raw}} - \bm{z}^*$. Evaluating the components yields $d_1 = \bm{u}^\top \bm{d}$ and $\bm{d}_2 = \bm{d} - d_1 \bm{u}$. The gradient of the scalarization $g^{\text{PBI}}$ with respect to the objective space $\bm{f}$ evaluates cleanly to:
\begin{equation}
\nabla_{\bm{f}} g^{\text{PBI}} = \nabla_{\bm{f}} d_1 + \theta_p \nabla_{\bm{f}} d_2 = \bm{u} + \theta_p \frac{\bm{d}_2}{\|\bm{d}_2\|}.
\end{equation}
This instantaneous gradient defines the steepest-descent path in the objective space mapping directly back to an \emph{effective target direction}:
\begin{equation}
\bm{\lambda}_{\text{eff}} = \bm{u} + \theta_p \frac{\bm{d}_2}{\|\bm{d}_2\|}.
\end{equation}

Consequently, PBI exploration centered on the reference state $\bm{x}_{\text{ref}}$ is equivalent to solving a classical weighted-sum MOO employing the dynamically synthesized direction $\bm{\lambda}_{\text{eff}}$. For each subsequent parameterized warm-start round, the classical controller instantaneously synthesizes the single-objective Hamiltonian by merely realigning the base mixing proportions:
\begin{equation}
\bar{J}_{uv} = \sum_{i=1}^{k} \lambda_{\text{eff}, i}\, J^{(i)}_{uv}, \quad \bar{h}_v = \sum_{i=1}^{k} \lambda_{\text{eff}, i}\, h^{(i)}_v
\end{equation}
By directly rebinding these $R_{ZZ}$ and $R_Z$ rotation strengths within the hardware circuit layers, the QAOA systematically explores geometry-aware $d_2$ exclusion topologies. The heavy-lifting of geometry calculation is perfectly offloaded into classical inter-round calculations, guaranteeing that $H_C = \sum \bar{J}_{uv} Z_u Z_v + \sum \bar{h}_v Z_v$ never exceeds two-body physical density limits mapping cleanly to generic QAOA pipelines.
\end{document}